\documentclass[12pt]{article}
\usepackage{epsfig,amsfonts,amssymb}
\usepackage{hyperref}
\usepackage{cite}
\input epsf.sty
\usepackage{cite}

\def\ZZZ{{\hbox{ Z\kern-1.6mm Z}}}
\def\RRR{{\hbox{\hskip-.3mm R\kern-2.4mm R}}}
\def\CCC{{\hbox{ C\kern-2.0mm C}}}
\def\zzz{{\hbox{z\kern-1mm z}}}

\newcommand{\nn}{\nonumber \\}

\newcommand{\qeq}{{\hbox{=\kern-2.3mm ? \kern.5mm }}}
\renewcommand{\qeq}{=}

\newcommand{\eps}{\epsilon}

\newcommand{\ve}{\varepsilon}

\newcommand{\JJ}{{\cal J}}

\newcommand{\OO}{{\cal O}}

\newcommand{\wt}{\widetilde}
\newcommand{\wh}{\widehat}

\newcommand{\NN}{{\cal N}}

\newcommand{\be}{\begin{equation}}
\newcommand{\ee}{\end{equation}}
\newcommand{\ben}{\begin{eqnarray}\displaystyle}
\newcommand{\een}{\end{eqnarray}}

\newcommand{\refb}[1]{(\ref{#1})}
\newcommand{\p}{\partial}
\newcommand{\sectiono}[1]{\section{#1}\setcounter{equation}{0}}

\def\one{{\hbox{ 1\kern-.8mm l}}}
\def\zero{{\hbox{ 0\kern-1.5mm 0}}}

\begin{document}

\baselineskip 24pt

\begin{center}
{\Large \bf Supersymmetry, Localization and Quantum Entropy
Function}

\end{center}

\vskip .6cm
\medskip

\vspace*{4.0ex}

\baselineskip=18pt

\centerline{\large \rm  Nabamita Banerjee, Shamik Banerjee, Rajesh
Kumar Gupta,}
\centerline{\large \rm  Ipsita Mandal and Ashoke Sen}

\vspace*{4.0ex}

\centerline{\large \it Harish-Chandra Research Institute}

\centerline{\large \it  Chhatnag Road, Jhusi,
Allahabad 211019, INDIA}

\vspace*{1.0ex}
\centerline{E-mail:  nabamita, bshamik, rajesh, ipsita, sen at hri.res.in}

\vspace*{5.0ex}

\centerline{\bf Abstract} \bigskip

$AdS_2/CFT_1$ correspondence leads to a
prescription for computing the degeneracy of black hole states
in terms of path integral over string fields living on the near horizon
geometry of the black hole. In this paper we make use of the
enhanced supersymmetries of the near horizon geometry and
localization techniques to argue that the path integral receives
contribution only from a special class of string field configurations
which are invariant under a subgroup of the supersymmetry 
transformations. We identify saddle points which are invariant
under this subgroup. 
We also use our analysis to show that the integration over infinite
number of zero modes generated by the asymptotic symmetries
of $AdS_2$ generate a finite contribution to the path
integral. 

\vfill \eject

\baselineskip=18pt

\tableofcontents

\sectiono{Introduction} \label{sintro}

Supersymmetric extremal black holes typically have an $AdS_2$ factor
in their near horizon geometry. Based on $AdS_2/CFT_1$ correspondence
refs.\cite{0809.3304,0810.3472,0903.1477}
proposed  a definite relation between the 
degeneracy $d_{hor}$ 
associated with the black hole horizon
and the partition function of string
theory on the near horizon geometry.  This relation takes the
form:
\be \label{ealt}
d_{hor} = \left\langle \exp\left[-
i  q_i\ointop d\theta \, A^{(i)}_\theta\right]
\right\rangle^{finite}_{AdS_2}\, ,
\ee
where $\langle ~\rangle_{AdS_2}$ 
denotes the unnormalized path integral
over all the fields in string theory,
 weighted by $e^{-A}$ where $A$ is the Euclidean
action, with the  boundary
condition that asymptotically the field configuration approaches
the near horizon geometry of the black hole containing an
$AdS_2$ factor. $\{A^{(i)}\}$ denote the set of all $U(1)$
gauge fields of string theory
living on the $AdS_2$ component of the near
horizon geometry, $q_i$ is the $i$-th
electric charge carried by the
black hole
and $\ointop \, d\theta\, A^{(i)}_\theta$
denotes the integral of the $i$-th gauge field along the boundary
of $AdS_2$. $q_i$ in particular also includes the angular momentum
of the black hole\cite{0606244}.
The superscript `${finite}$' refers to the finite part of the
amplitude defined as follows.
If we represent $AdS_2$ as the Poincare disk,
regularize the infinite volume of $AdS_2$ by putting an
infrared cut-off and denote by
$L$ the length of the boundary of this regulated
$AdS_2$, then for large cutoff, \i.e.\ large $L$,
the amplitude 
has the form $e^{C L + \OO(L^{-1})}\times\Delta$ 
where $C$ and $\Delta$ are $L$-independent
constants. The finite part of the amplitude
is defined to be the constant $\Delta$, and has been
named the quantum entropy function.

In computing the path integral
we need to keep the asymptotic values of electric fields
fixed\footnote{These fixed values are determined in terms of
$q_i$ by requiring that the boundary terms in the variation of
the action cancel.}
and integrate over
the constant modes of the gauge fields. 
As a result the path integral directly computes 
the degeneracy in a fixed charge sector,
\i.e.\ entropy in the microcanonical ensemble, as opposed to
a partition function. 
Due to the same reason it also computes the degeneracy in
a fixed angular momentum sector. This allows us to compute the
index in terms of the degeneracy
$d_{hor}$ which can then be compared with
the result for the microscopic degeneracy\cite{0903.1477}.
It was found in 
\cite{0809.3304,0805.0095,0806.0053} that
in the semiclassical limit the result of this
path integral reproduces correctly the exponential of the Wald
entropy\cite{9307038,9312023,9403028,9502009} 
via the entropy function formalism\cite{0506177,0708.1270}
even after taking into account higher derivative corrections to the
classical action.

In four space-time dimensions
supersymmetry requires the black holes to be spherically symmetric,
and as a consequence the near horizon geometry has an $AdS_2\times
S^2$ factor. 
For $1/8$ BPS black holes in $\NN=8$ supersymmetric
theories, $1/4$ BPS black holes in $\NN=4$ supersymmetric theories
and $1/2$ BPS black holes in $\NN=2$ supersymmetric theories,
the $SL(2,\RRR)\times SO(3)$ isometry of the near horizon geometry
gets enhanced to the $SU(1,1|2)$ supergroup. The goal of this paper
will be to simplify the path integral over string fields 
appearing in the definition of
\refb{ealt} by making use of these isometries. In
particular we shall use localization 
techniques\cite{heckman,wittens,witten92,
9112056,9204083,9511112,zaboronsky,0206161,0712.2824,ati1,ati2}
to show that  the
path integral receives contribution only from a special class
of field configurations which preserve a
particular subgroup of 
$SU(1,1|2)$. 

The effect of localization in these
theories have been studied earlier in \cite{0608021} 
for analyzing the 
world-sheet instanton contributions. Our goal is quite different as we
use it to classify string field configurations which could contribute
to the path integral. Thus for example our analysis can be used to
restrict the
saddle points, \i.e.\ classical string
field configurations, which contribute to the 
path integral over string fields. These 
are relevant
for computing non-perturbative corrections to the 
quantum entropy function 
whereas the world-sheet instanton corrections
are relevant only for
string loop
computation.

The rest of the paper is organised as follows. In \S\ref{ssymm} we
describe the algebra underlying the $SU(1,1|2)$ group and also
the reality condition on the various generators required to represent
the symmetries of the Euclidean near horizon geometry.
In \S\ref{slocal} we use localization techniques
developed in \cite{9511112,zaboronsky} to argue that the path
integral receives contribution only from a special class of string
field configurations invariant under a special subgroup $H_1$ of the
$SU(1,1|2)$ group. In \S\ref{sorbit} we use the results of
\S\ref{slocal} to show that integration over the bosonic and fermion
zero modes, generated by an infinte dimensional group of asymptotic
symmetries, actually gives a finite result to the path integral. 
In \S\ref{sexam} we
give some examples of $H_1$-invariant saddle points which 
contribute to the path integral. In \S\ref{scomm} we discuss possible
application of our result to further simplify the analysis of quantum
entropy function and also a possible application to computing the
expectation values of  circular 't Hooft - Wilson loop operators in
superconformal gauge theories following \cite{0904.4486}. 
In appendix 
\ref{killingspinor} we analyze the killing spinors in the near
horizon geometry of a specific class of quarter BPS black holes in
type IIB string theory compactified on $K3\times T^2$ and show that
they indeed generate the $su(1,1|2)$ algebra described in
\S\ref{ssymm}. 

\sectiono{Symmetries of Euclidean $AdS_2\times S^2$} \label{ssymm}

We begin by writing down the global part of
the $\NN=4$ superconformal algebra 
in 1+1 dimensions.
Its non-vanishing commutators are
\ben\label{esp1}
&& [L_m, L_n] = (m-n) \, L_{m+n}\, , \nn
&& [J^3 , J^\pm ] = \pm J^\pm \, , 
\qquad [J^+ , J^- ] =  2\, J^3 \, , \nn
&& [L_n, G^{\alpha\pm}_r] = \left({n\over 2}-r\right) 
G^{\alpha\pm}_{r+n}
\, , \nn
&& [J^3 , G^{\alpha\pm}_r] = \pm\, {1\over 2}\, 
G^{\alpha\pm}_{r }\, , 
\quad 
[J^\pm , G^{\alpha\mp}_r] = G^{\alpha\pm}_r  \, ,\nn
&& \{G^{+\alpha}_r, G^{-\beta}_s\} = 2\, \eps^{\alpha\beta}\,
L_{r+s} - 2\, (r-s)\, 
(\eps\sigma^i)_{\beta\alpha}\, J^i\nonumber \\
&& \eps^{+-}=-\eps^{-+}=1, \quad \eps^{++}=\eps^{--}=0
\, , \qquad m,n=0,\pm 1, \quad r,s=\pm{1\over 2},
 \quad \alpha,\beta=\pm\, .\nn
\een
The right superscript of $G_r$ 
denotes the transformation properties under
the SU(2) current algebra whose zero modes are denoted by
$(J^3,J^\pm=J^1\pm i J^2)$. There is also an $SU(2)$ group
acting on the left superscript. This describes an outer automorphism
of the supersymmetry algebra but is not in general
a symmetry of the theory.
In eq.\refb{esp1} and
in the rest of this paper we shall use the
convention that when an equation involves $\pm$ or $\mp$, 
it represents only two equations -- first one being obtained by choosing
the upper sign in all the terms and the second one obtained by
choosing the lower sign in all the terms.
The supergroup generated by the algebra \refb{esp1} is
known as $SU(1,1|2)$. 

In the above, the action of the Virasoro generators on the coordinate
$u$ labelling the upper half plane (UHP) is of the form
\be \label{esrep1}
L_n = -u^{n+1}\p_u - \bar u^{n+1}\p_{\bar u}\, .
\ee
However
while describing symmetries of the Euclidean $AdS_2\times S^2$,
which is isomorphic to $UHP\times S^2$ it is more natural to 
use the Virasoro generators
\be \label{esrep2}
L_n = -\left(i\, u^{n+1}\p_u + i\, \bar u^{n+1}\p_{\bar u}\right)\, ,
\ee
so that the elements of SL(2,\RRR) can be labelled as
$\exp(is_n L_n)$ with real parameters $s_n$,
just as $\exp(it_i J^i)$ labels an element of the $SU(2)$ group
for real $t_i$.
The corresponding algebra is obtained from \refb{esp1} by
scaling the
Virasoro generators by $i$.  For later convenience we shall also
multiply $G^{+\alpha}_r$ by $e^{i s_0}$ and
$G^{-\alpha}_r$ by $i \, e^{-i s_0}$ for some arbitrary fixed
phase $e^{is_0}$. This gives\footnote{Note that while computing
the commutators we regard the action of the generators as active
transformation.}
\ben\label{ekx1}
&& [L_m, L_n] = i\, (m-n) \, L_{m+n}\, , \nn
&& [J^3, J^\pm] = \pm J^\pm \, , 
\qquad [J^+, J^-] =  2\, J^3 \, , \nn
&& [L_n, G^{\alpha\pm}_r] = i\, \left({n\over 2}-r\right) 
G^{\alpha\pm}_{r+n}
\, , \nn
&& [J^3, G^{\alpha\pm}_r] = \pm\, {1\over 2}\, 
G^{\alpha\pm}_{r}\, , 
\quad 
[J^\pm , G^{\alpha\mp}_r] = G^{\alpha\pm}_r  \, ,\nn
&&  \{G^{+\alpha}_r, G^{-\beta}_s\} = 2\, \eps^{\alpha\beta}\,
L_{r+s} - 2\, i\, (r-s)\, 
(\eps\sigma^i)_{\beta\alpha}\, J^i\nonumber \\
&&
\eps^{+-}=-\eps^{-+}=1, \quad \eps^{++}=\eps^{--}=0
\, ,  \qquad m,n=0,\pm 1, \quad r,s=\pm{1\over 2},
 \quad \alpha,\beta=\pm\, .\nn
\een

Often it is convenient to represent $AdS_2$ as a unit disk labelled by
a
coordinate $w$ related to $u$ via:
\be \label{ekx3}
w = {1+i\, u\over 1 - i\, u}\, .
\ee
In the $w$ coordinate system
\be \label{ekx4}
L_n = {i\over 2} \, \left[i^{n}\,
(1+w)^{1-n} \, (1-w)^{1+n}\, \p_w + c.c.\right]\, .
\ee
On the other hand the action of the $J^i$'s 
on the stereographic  coordinate
$z$ of the sphere $S^2$ takes the form
\be \label{ejj1}
J^3 = (z\p_z-\bar z\p_{\bar z}) , \qquad J^+ = z^2\p_z
+\p_{\bar z}, \qquad J^- =   -\bar z^2 \p_{\bar z} - \p_{z}
\, .
\ee
It is easy to see that the $AdS_2\times S^2$ metric
\be \label{ekx5}
ds^2 = 4\, v \, {dw\, d\bar w\over (1-\bar w w)^2}
+ 4\, u\, {dz\, d\bar z\over (1+\bar z z)^2} \, ,
\ee
where $u$ and $v$ are constants, 
is invariant under these transformations. 
Making the coordinate transformations
\be \label{ectrswz}
w=\tanh{\eta\over 2}\, e^{i\theta}, \qquad z = \tan{\psi\over 2}
e^{i\phi}\, ,
\ee
we can express the metric  \refb{ekx5} as
\be \label{ewzdef}
ds^2 = v(d\eta^2 +\sinh^2\eta\, d\theta^2) +
u(d\psi^2 + \sin^2\psi \, d\phi^2)\, .
\ee
This form of the $AdS_2\times S^2$ 
metric coincides with the one given in
\refb{eb1}. In appendix
\ref{killingspinor} we have shown 
that the full $SU(1,1|2)$ group
generated by
$\wh L_0$, $\wh L_\pm$, $J^3$, $J^\pm$ and 
$\wh G^{\alpha\beta}_\gamma$ describes a symmetry of the near
horizon $AdS_2\times S^2$ geometry of Euclidean 
BPS black holes.

We now define
\be \label{ekx6}
\wh L_0 = {1\over 2}\, (L_1+L_{-1})
\, ,  \quad 
\wh L_\pm = L_0 \pm {i\over 2} (L_1 - L_{-1})\, ,
\quad \wh G^{\alpha\beta}_\pm = G^{\alpha\beta}_{1/2} \mp 
i\, G^{\alpha\beta}_{-1/2}\, .
\ee
{}From eqs.\refb{ekx4}, \refb{ekx6} we see that the action of
$\wh L_0$, $\wh L_\pm$ on the $w$-plane is given by
\be \label{ekx8}
\wh L_0 = (w\p_w -\bar w\p_{\bar w}), \quad
\wh L_+ = -i (w^2\p_w -\p_{\bar w}), \quad \wh L_-
= i ( \p_w - \bar w^2 \p_{\bar w})\, .
\ee
This shows that $\wh L_0$ has the 
interpretation of the generator
of rotation about the origin in the $w$-plane.
In terms of these new generators 
the non-vanishing (anti-)commutators of the
$su(1,1|2)$ algebra take the form
\ben\label{ekx7}
&& [\wh L_0, \wh L_\pm] = \pm \, \wh L_\pm\, , \qquad
[\wh L_+, \wh L_-] = -2\, \wh L_0\, , \nn
&& [J^3, J^\pm] = \pm J^\pm \, , 
\qquad [J^+, J^-] = 2\, J^3 \, , \nn
&& [\wh L_0, \wh G^{\alpha\beta}_\pm] = \pm\, 
{1\over 2}\, \wh G^{\alpha\beta}_\pm
\, ,  \quad [\wh L_\pm, \wh G^{\alpha\beta}_\mp]= -i\,
\wh G^{\alpha\beta}_\pm\, , \nn
&& [J^3, \wh G^{\alpha \pm}_\beta] = \pm\, {1\over 2}\, 
\wh G^{\alpha\pm}_{\beta}\, , 
\quad 
[J^\pm , \wh G^{\alpha\mp}_\beta] = \wh G^{\alpha\pm}_\beta  
\, ,\nn
&& \{\wh G^{+\alpha}_\pm, \wh G^{-\beta}_\mp\}
= 4\, \eps^{\alpha\beta}\, \wh L_0 \pm 4\, (\eps\sigma^i)_{\beta\alpha}
\, J^i\, , \qquad \{\wh G^{+\alpha}_\pm, \wh G^{-\beta}_\pm\}
=\mp\, 4\, i\, \eps^{\alpha\beta}\, \wh L_\pm\, . 
\een

Note that an element of the form
$\exp\left[i(\xi^0 \wh L_0+\xi^+\wh L_+ + \xi^- \wh L_-
+ \eta_3 J^3 + \eta_+ J^++\eta_- J^-)\right]$ will be an element
of the $SL(2,\RRR)\times SU(2)$ group if we have
\be \label{erealboson}
(\xi^0)^*=\xi^0, \quad (\xi^\pm)^*=\xi^\mp, \quad
(\eta_3)^*=\eta_3, \quad (\eta_\pm)^*=\eta_\mp\, .
\ee
We shall call these the reality conditions on the bosonic generators.
We shall now impose a similar reality condition on the
fermionic generators, \i.e.\ specify the condition on the 
complex grassman
parameters $\theta^{\gamma}_{\alpha\beta}$ under which
$\exp\left[i \theta^\gamma_{\alpha\beta} G^{\alpha\beta}_\gamma
\right]$ describes an element of the $SU(1,1|2)$ group.
Any such rule must be compatible with the requirement that
if $\exp(iT_1)$ and $\exp(iT_2)$ are two elements of the
$SU(1,1|2)$ group, then $\exp([T_1,T_2])$ must also be an
element of this group. The following constraint on
$\theta^\gamma_{\alpha\beta}$ is compatible with
this rule:\footnote{We need to remember that if $\theta_1$ and
$\theta_2$ are real grassman parameters then $(\theta_1\theta_2)^*
=\theta_2\theta_1=-\theta_1\theta_2$. \label{f1}}
\be \label{ethetarule}
\left(\theta_{\alpha\beta}^\gamma\right)^* 
= \eps^{\alpha\alpha'}\eps^{\beta\beta'}\, 
\theta_{\alpha'\beta'}^{-\gamma}\, .
\ee
Equivalently we can say that
\be \label{eequiv}
\exp\left[i\theta\left(\wh G^{\alpha\beta}_\gamma
+ \eps^{\alpha\alpha'}\eps^{\beta\beta'}  
\wh G^{\alpha'\beta'}_{-\gamma}\right)\right] 
\quad \hbox{and}\quad
\exp\left[\theta\left(\wh G^{\alpha\beta}_\gamma
- \eps^{\alpha\alpha'}\eps^{\beta\beta'}  
\wh G^{\alpha'\beta'}_{-\gamma}\right) \right]\, ,
\ee
are elements of $SU(1,1|2)$ for real $\theta$.
We shall proceed with this
choice. If we now define
\ben \label{ehg1}
&& Q_1 = \wh G^{++}_+ + \wh G^{--}_-, \qquad Q_2 = -i \left(
\wh G^{++}_+ - \wh G^{--}_-\right)\, , \nn
&& Q_3 = -i\left( \wh G^{-+}_+ + \wh G^{+-}_-\right), 
\qquad Q_4 = 
\wh G^{-+}_+ - \wh G^{+-}_- \, , \nn
&& \wt Q_1 = \wh G^{++}_- + \wh G^{--}_+, \qquad \wt
Q_2 = -i \left(
\wh G^{++}_- - \wh G^{--}_+\right)\, , \nn
&& \wt Q_3 = -i\left( \wh G^{-+}_- + \wh G^{+-}_+\right), 
\qquad \wt Q_4 = 
\wh G^{-+}_- - \wh G^{+-}_+ \, ,
\een
then $\exp(i\theta Q_i)$ and $\exp(i\theta\wt Q_i)$ are 
elements of $SU(1,1|2)$  for real
$\theta$.
In that case we have
\be \label{ehg2}
\{Q_i,Q_j\} = 8 \, \delta_{ij}\,
(\wh L_0 - J^3), \quad \{\wt Q_i, \wt Q_j\}
= 8  \, \delta_{ij}\,
(\wh L_0 + J^3), \quad [\wh L_0-J^3, Q_i]=0, \quad
[\wh L_0 + J^3, \wt Q_i] = 0\, .
\ee
Besides this,
$\{Q_i,\wt Q_j\}$ are given by
linear combinations of $J^\pm$ and $\wh L_\pm$, $[\wh L_0-J^3,
\wt Q_i]$ are given by linear combinations of $\wt Q_i$,
$[\wh L_0+J^3,Q_i]$ are given by linear combinations
of $Q_i$, $[J^\pm,Q_i]$ and  $[\wh L_{\pm},Q_i]$ 
are given by linear combinations of $\wt Q_i$ and
$[J^\pm,\wt Q_i]$, $[\wh L_{\pm},
\wt Q_i]$ are given by linear combinations of $Q_i$.
Precise form of these relations can be determined from
\refb{ekx7} and \refb{ehg1}, but we shall not write them down
explicitly.

Given the reality condition on the various generators, 
we can label an element of $SU(1,1|2)$ as 
\ben \label{egroup}
g(\xi,\bar\xi,\eta,\bar\eta,\sigma,\wt\sigma,\{\theta_{\alpha\beta}\},
\{\chi_i\})
&=& \exp\left[i\left\{\bar \xi \wh L_+ + \xi \wh L_-+\bar\eta J^++\eta J^-
+\theta_{\alpha  +}\wh G^{\alpha +}_- + \theta_{\alpha -} 
\wh G^{\alpha -}_+
\right\}\right] \nonumber \\ &&
\times
\exp\left[i \sigma \, (\wh L_0 + J^3) \right]
 \times \exp\left[ i\left\{\sum_{k=1}^4
\chi_k Q_k + \wt\sigma (\wh L_0 - J^3)\right\}\right]\, , \nn
\een
where $\xi$, $\eta$ are complex bosonic parameters, $\sigma$, $\wt\sigma$
are real bosonic parameters, $\chi_i$ are real grassman parameters
and $\theta_{\alpha\beta}$ are complex grassman parameters satisfying
the reality condition
\be \label{ethetareal}
(\theta_{\alpha\beta})^* = \eps^{\alpha\alpha'}\eps^{\beta\beta'}
\theta_{\alpha'\beta'}\, .
\ee
Let us also denote by $H_0$
the subgroup of
$SU(1,1|2)$ generated by
\be \label{ekx9}
\wh L_0- J^3, \quad Q_1, \quad Q_2, \quad Q_3, \quad
Q_4\, .
\ee
The non-vanishing (anti-)commutators of $H_0$ are
\be \label{esubal}
\left[\wh L_0- J^3, Q_i\right] =0, \qquad \{Q_i, Q_j\}
= 8\, \delta_{ij}\, 
(\wh L_0 - J^3)\, .
\ee
Then in \refb{egroup}
the parameters  $\wt\sigma$ and $\{\chi_i\}$ 
parametrize
an element of $H_0$ and the parameters $\sigma$, $\xi$, 
$\eta$ and $\theta_{\alpha\beta}$ parametrize the coset
$G/H_0$.

Finally another subgroup of $SU(1,1|2)$ (and $H_0$) that will 
play an important role in our analysis is the subgroup $H_1$
generated by $Q_1$ and $(\wh L_0-J^3)$.

\sectiono{Localization} \label{slocal}

In computing the quantum entropy function, -- the partition function
of string theory on the near horizon geometry of the black hole --
we need to integrate over all string field configurations. 
In order to carry out the path integral, which involves
integration over infinite number of modes, it will be
useful to fix the order in which we carry out the
integration. We shall adopt the following definition of the
path integral: first we shall integrate over the orbits of the
subgroup $H_1$ generated by $Q_1$ and $(\wh L_0-J^3)$,
then over the orbits generated by the others $Q_i$'s belonging to the
subgroup $H_0$ and
then carry out the integration over the remaining variables in
some order. As we shall see this definition will allow us to arrive
at simple results on which configurations could contribute
to the path integral. Our approach follows closely that of
\cite{9511112}. Throughout this analysis we shall implicitly
assume that the theory admits a formalism in which at least
the $H_1$ subalgebra of the 
$su(1,1|2)$ algebra, generated by $Q_1$ and
$(\wh L_0-J^3)$,
is realized  off-shell. It may be possible to 
achieve this by generalizing
the trick used in \cite{0712.2824} for $\NN=4$ supersymmetric
gauge theories. Finally we shall ignore the various issues related
to gauge fixing. For  supersymmetric gauge theories in
four dimensions gauge fixing introduces various subtleties in the
proof of localization\cite{0712.2824}. 
However eventually these can be overcome, and
we shall assume that similar results will hold for supergravity as
well.

Formally the division of the path integral into orbits 
of $H_1$ and  directions transverse to these orbits
can be done by
manipulating the integral using Fadeev-Popov 
method.\footnote{Unlike in the case of a gauge symmetry here we
do not divide the path integral by the volume of the `gauge
group' $H_1$. The rest of the manipulation proceeds exactly as in the
case of a gauge theory.}
By expressing an element of $H_1$ as
\be \label{elh}
h=\exp(i\alpha Q_1 + i\beta (\wh L_0-J^3))
\ee
we can express the path integral as
\be \label{efp1}
\left[\int dh \right] \, \left[\int \, e^{-A}\, \left(\prod_a \delta(F^a)
\right) \, \left.
{\rm sdet}{\delta F^a_{\vec \tau}\over 
\p \tau^b}\right|_{\vec\tau=0}\right]\, ,\
\ee
where $\int dh$ denotes integration over the group $H_1$ with Haar
measure, $A$ is the Euclidean action,\footnote{We are
including in $A$ the bulk and the boundary contributions to the
action including the $i\ointop \vec q\cdot \vec A$ term that is
necessary to make the path integral well 
defined\cite{0809.3304,0903.1477}.
We shall also be implicitly assuming that the boundary terms have
been chosen so that all the supersymmetries of the bulk theory
are
preserved.}
$F^a$ are a pair of `gauge fixing functionals' of the field
configuration, $\vec\tau$ denote collectively the
 parameters $(\alpha, \beta)$ labelling the
elements of the group $H_1$ and $F^a_{\vec\tau}$ is the transform of 
$F^a$ by the group
element corresponding to the parameters $\vec\tau$.
We now note that the integration over $H_1$ has a bosonic direction
$\beta$ which parametrizes a compact $U(1)$ group and hence
gives a finite result, and a fermionic direction
$\alpha$. By the standard rules of integration over grassman
parameters the fermionic integral gives a zero, making the
whole integral vanish.

This argument breaks down around a configuration $\Phi$ which
is 
invariant under a subgroup of $H_1$,
since 
the matrix $(\delta F^a_{\vec \tau}/\delta \tau^b)$ in
\refb{efp1} becomes
degenerate at this point. 
In this case we proceed as follows. 
First of all note that 
a subgroup of $H_1$ can either be the whole of $H_1$ or
the $U(1)$ group generated by $(\wh L_0-J^3)$.
However if $\Phi$ is invariant only 
under $(\wh L_0-J^3)$, then the
zero eigenvector of the matrix $\left. 
{\delta F^a_{\vec \tau}/
\p \tau^b}\right|_{\vec\tau=0}$ is along the bosonic direction
corresponding to the $U(1)$ transformation. This makes the
${\rm sdet}$ factor in \refb{efp1} vanish on the configuration $\Phi$
but does not generate any divergence in the integrand. Hence our
earlier argument can still be applied to show that the
$\int dh$  factor makes the integral vanish. Thus 
the configuration $\Phi$ must be invariant under 
both $Q_1$ and $(\wh L_0-J^3)$.
This allows us to 
choose the coordinates of the configuration space,
measuring fluctuations around the
configuration $\Phi$, as follows. 
First by Fourier decomposing these fluctuations in the
$(\theta -\phi)$ coordinates we can choose them to be eigenvectors
of $(\wh L_0-J^3)$ with definite eigenvalues $m\in \ZZZ$. For
example for a scalar field a deformation of the form 
$e^{i m (\theta-\phi)/2}\, f(\theta+\phi,r,\psi)$ 
for any arbitrary function $f$
will have this property. Let us parametrize
the set of all such bosonic fluctuations by coordinates $z_{(m)}^s$.
The complex conjugate deformation, labelled by $z_{(m)}^{s*}$ will
have $(\wh L_0-J^3)$ eigenvalue $-m$. To avoid double
counting we shall denote the fluctuations
with positive $m$ by $z_{(m)}^s$ and fluctuations with negative
$m$ by $z_{(m)}^{s*}$. As $s$ runs over different values,
the parameters $z_{(m)}^s$ produce the complete set of
bosonic deformations with $(\wh L_0-J^3)$ eigenvalue $m$.
Now for $m\ne 0$, the action of the generator $Q_1$ on such a
bosonic deformation cannot vanish since $Q_1^2=4(\wh L_0-J^3)$
acting on the fluctuation does not vanish. Instead this will generate
a particular fermionic deformation with $(\wh L_0-J^3)$ eigenvalue
$m$.
Let us denote the parameter associated with the fermionic
deformation by $\zeta_{(m)}^s$. 
Finally we shall call the $m=0$ bosonic and fermionic modes
collectively as $\vec y$. Since the original configuration
$\Phi$ is the origin of the coordinate system, all the coordinates
vanish at $\Phi$.
We can now
write\footnote{Our convention for defining the action of $Q_1$
on the parameters will be as follows. Take a general field
configuration labelled by $(\{ z^s_{(m)}\}, \{\zeta^s_{(m)}\},
\vec y)$ and act on it by the transformation
$(1+i\eps Q_1)$. The new configuration can be associated with
a new set of values of the various parameters. We call the
parameters associated with the new configuration as
$(\{ z^s_{(m)} + i\eps Q_1\, z^s_{(m)} \}, \{\zeta^s_{(m)}
+ i\eps Q_1\, \zeta^s_{(m)}\},
\vec y+ i\eps Q_1\, \vec y)$.}
\be \label{esk1}
Q_1 \, z_{(m)}^s= \zeta_{(m)}^s, \quad Q_1 \, \zeta_{(m)}^s
= 4m \, z_{(m)}^s\, ,
\ee
where the second equation follows from the fact that $(Q_1)^2
\, z_{(m)}^s = 4\, m\, z_{(m)}^s$. Using the reality of the operator
$(i\eps Q_1)$ and the rules for complex conjugation of grassman
variables described in footnote \ref{f1}, the complex conjugate
relations of \refb{esk1} can be expressed in the
form\footnote{To see this we can write $z_{(m)}^s
=z_{(m)R}^s+i z_{(m)I}^s$, $\zeta_{(m)}^s=
\zeta_{(m)R}^s + i \zeta_{(m)I}^s$ with real 
$z_{(m)R}^s$, $z_{(m)I}^s$, $\zeta_{(m)R}^s$ and
$\zeta_{(m)I}^s$, and then compare the real and imaginary parts
of \refb{esk1} after multiplying both sides by $i\theta$, 
keeping in mind that the operator 
$i\theta Q$ for real grassman parameter $\theta$ takes a real
variable to a real variable, and also that given two real
grassman variables $\theta_1$, $\theta_2$, $\theta_1\theta_2$
is imaginary. Eq.\refb{esk2} follows from this immediately.
\label{fsee}}
\be \label{esk2}
Q_1 \, z_{(m)}^{s*}= \zeta_{(m)}^{s*}, \quad Q_1 \, 
\zeta_{(m)}^{s*}
= -4m \, z_{(m)}^{s*}\, .
\ee
$\zeta_{(m)}^s$ for different values of $s$ give the complete set of
fermionic deformations with $(\wh L_0-J^3)$ eigenvalue
$m$ and $\zeta_{(m)}^{s*}$ for different values 
of $s$ give the complete set of
fermionic deformations with $(\wh L_0-J^3)$ eigenvalue
$-m$. To see this let us assume the contrary, \i.e.\ that there is
a fermionic coordinate $\chi_{(m)}$ carrying $\wh L_0 - J^3$
eigenvalue $m$ that is linearly independent of the 
$\zeta^s_{(m)}$'s (up to quadratic and higher powers of the
other coordinates). Since the origin is $Q_1$ invariant,
$Q_1\chi_{(m)}$ must vanish at the origin. On the other hand
if $Q_1\chi_{(m)}$ is bilinear in the coordinates
$(\{z_{(m)}^s\}, \{z_{(m)}^{s*}\}, \{\zeta_{(m)}^{s}\},
\{\zeta_{(m)}^{s*}\}, \chi_{(m)}, \vec y)$ then it will be
impossible to satisfy the $Q_1^2 \chi_{(m)}=4m\chi_{(m)}$
condition since the action of $Q_1$ on each of the coordinates
produces a term linear and higher order in these coordinates.
Thus $Q_1\chi_{(m)}$ must be a linear combinations of the
complete set of bosonic coordinates $\{z^s_{(m)}\}$ carrying
$\wh L_0-J^3$ eigenvalue $m$ up to additional higher order terms
in the coordinates. Applying $Q_1$ on either side we see that
$\chi_{(m)}$ must be a linear combination of the coordinates
$\zeta_{(m)}^s$ up to additional higher order terms, in contrary to
our original assumption that $\chi_{(m)}$ is linearly independent
of the other $\zeta_{(m)}^s$'s.

The coordinates  
$(\{z_{(m)}^s\}, \{z_{(m)}^{s*}\}, \{\zeta_{(m)}^{s}\},
\{\zeta_{(m)}^{s*}\})$
will in particular include the deformations generated
by the elements of   $SU(1,1|2)$ outside the subgroup 
generated by the $Q_i$'s and $(\wh L_0\pm J^3)$, since
such deformations will carry non-zero $\wh L_0 - J^3$ charge.
If for example we use the parametrization given in 
\refb{egroup} for an element of $SU(1,1|2)$, then the parameters
$\xi$, $\bar\eta$ and $\theta_{\alpha +}$ will carry $(\wh L_0
-J^3)$ eigenvalue $+1$, and their complex conjugate parameters
will carry $(\wh L_0
-J^3)$ eigenvalue $-1$. 

Now the path integral over the various fields can be regarded as
integral over the parameters $z_{(m)}^s$, $z_{(m)}^{s*}$,
$\zeta_{(m)}^s$ and $\zeta_{(m)}^{s*}$ for different values of
$s$ and $m\ne 0$ together with integration over
the variables $\vec y$.
Thus we have an integral
\be \label{esk3}
I = \int d\vec y\, \prod_{m>0, s} dz_{(m)}^s\, dz_{(m)}^{s*}\,
d\zeta_{(m)}^s\, d\zeta_{(m)}^{s*} \, \JJ\, e^{-A}\, .
\ee
where $\JJ$ represents 
any measure factor which might arise from changing the
integration variables
to $(\vec y, \vec z, \vec z^*,
\vec \zeta,\vec \zeta^*)$. We now deform this to another
integral
\be \label{esk4}
I(t) = \int d\vec y\, \prod_{m>0} dz_{(m)}^s\, dz_{(m)}^{s*}\,
d\zeta_{(m)}^s\, d\zeta_{(m)}^{s*} \, \JJ\, 
e^{-A -t\, Q_1\, F}\, ,
\ee
where $t$ is a positive real parameter and
\be \label{esk5}
F = \sum_{m>0} \sum_s \, z_{(m)}^{s*}\, \zeta_{(m)}^s\, .
\ee
This gives
\be \label{esk5a}
Q_1 \, F = \sum_{m>0} \sum_s\,  \left[ 4m\, z_{(m)}^{s*}\, z_{(m)}^s
+\zeta_{(m)}^{s*} \, \zeta_{(m)}^s\right]\, . \ee
Furthermore, since by construction $F$ is 
invariant under $(\wh L_0-J^3)$, we have
\be \label{esk6}
Q_1^2 F = 0\, .
\ee
This equation, together with the supersymmetry invariance of the
action ($Q_1A=0$) can be used to get
\ben \label{esk7}
\p_t I(t) &=& \int d\vec y\, \prod_{m>0} dz_{(m)}^s\, dz_{(m)}^{s*}\,
d\zeta_{(m)}^s\, d\zeta_{(m)}^{s*} \, \JJ\,
(-Q_1F) \, e^{-A -t\, Q_1\, F}
\nn 
&=& -\int d\vec y\, \prod_{m>0} dz_{(m)}^s\, dz_{(m)}^{s*}\,
d\zeta_{(m)}^s\, d\zeta_{(m)}^{s*} \, \JJ\, 
Q_1\left( F\, e^{-A -t\, Q_1\, F}
\right) = 0\, ,
\een
where in the last step we have used $Q_1$ invariance of the
path integral measure. Thus $I(t)$ is independent of $t$, and
has the same value in the limits $t\to 0$ and $t\to\infty$. 
Noting that
in the $t\to 0$ limit $I(t)$ reduces to $I$, and using 
\refb{esk4}, \refb{esk5a} we get
\be \label{esk8}
I=\lim_{t\to\infty} 
\int d\vec y\, \prod_{m>0} dz_{(m)}^s\, dz_{(m)}^{s*}\,
d\zeta_{(m)}^s\, d\zeta_{(m)}^{s*} \, \JJ\, e^{-A -t\,
\sum_{m>0} \sum_s\,  \left[ 4\, m\, z_{(m)}^{s*}\, z_{(m)}^s
+\zeta_{(m)}^{s*} \, \zeta_{(m)}^s\right]}\, .
\ee
In the $t\to\infty$ limit the $z_{(m)}^s$ and $\zeta_{(m)}^s$
dependent terms inside the action $A$ are subleading. Thus
up to an overall $t$ independent normalization
constant,\footnote{This normalization constant can of course
be absorbed into a redefinition of the measure $\JJ$. 
Alternatively, we
could define $\zeta_{(m)}^s$ with a different normalization so that
eqs.\refb{esk1} take the form $Q_1 \, z_{(m)}^s= \alpha_m
\zeta_{(m)}^s, \quad Q_1 \, \zeta_{(m)}^s
= 4m \,\alpha_m^{-1}\, z_{(m)}^s$ for some constant $\alpha_m$.
By adjusting $\alpha_m$ we could ensure that the replacement
of the $t$ dependent exponential factor by \refb{esk9} does
not require any additional normalization.}
the
$e^{-t\,
\sum_{m>0} \sum_s\,  \left[ 4\, m\, z_{(m)}^{s*}\, z_{(m)}^s
+\zeta_{(m)}^{s*} \, \zeta_{(m)}^s\right]}$ term in the
$t\to\infty$ limit is equivalent
to inserting in the path integral a factor of
\be\label{esk9}
\prod_{m>0} \prod_s\, \delta\left(z_{(m)}^s\right)
\delta\left(z_{(m)}^{s*}\right)
\delta(\zeta_{(m)}^{s*})\delta(\zeta_{(m)}^s)\, .
\ee
This shows that the path integral is localized in the subspace of 
$(\wh L_0-J^3)$ invariant deformations parametrized by
the coordinates $\vec y$. In particular it restricts integration
over the orbits of $SU(1,1|2)$, generated by the action of
\refb{egroup} on any $(\wh L_0-J^3)$ 
invariant configuration, to the
subspace
\be \label{esubres}
\xi=0, \quad \eta=0, \quad \theta_{\alpha\beta}=0\, .
\ee 
More generally, since $\wh L_0$ and $J^3$
generate translations along $\theta$ and $\phi$ directions of
$AdS_2\times S^2$ respectively, restriction to $\wh L_0-J^3$
invariant subspace amounts to restricting the path integral over
field configurations which depend on $\theta$ and $\phi$ only
through the combination $(\theta+\phi)$.

We can further localize the $\vec y$ integral onto $Q_1$-invariant
subspace. Intuitively this can be understood by noting 
that unless $\vec y$
is invariant under $Q_1$, the orbit
of $Q_1$ through a point $\vec y$ will give a vanishing contribution
to the integral\cite{witten92,zaboronsky}. 
Thus the contribution to the integral must come from the
$Q_1$ invariant subspace of the $(\wh L_0-J^3)=0$ subspace.
Formally this can be established as follows. Let $(\{\vec w^\alpha\}
,\{\zeta^a\})$
denote the bosonic and fermionic components of $\vec y$. Then
we can write
\be \label{eztrs}
Q_1\, \zeta^a = f^a(\vec w, \vec \zeta)\, ,
\ee
for some functions $f^a$. We now insert into the path
integral a term
\be \label{eaddz}
\exp\left[-t Q_1 \, \sum_a \, \zeta^a\, f^a(\vec w, \vec \zeta )\right]
= \exp\left[-t \, \, \sum_a \, f^a(\vec w, \vec \zeta )\, 
f^a(\vec w, \vec \zeta )\right]\, .
\ee
Nilpotence of $Q_1$ and $Q_1$ invariance of the original action
can be used to argue that
the path integral is independent of $t$. Restriction of the path integral
to the purely bosonic subspace $\zeta^a=0$ now has a factor
$\exp\{-t \sum_a \, f^a(\vec w, \vec 0)\, 
f^a(\vec w, \vec 0)\}$. Thus in the $t\to \infty$ limit the path integral
is restricted to the subspace $f^a(\vec w, \vec 0)=0$ in the
$\vec\zeta=0$ sector. This is precisely the $Q_1$ invariant subspace
of purely bosonic configurations.

This establishes 
that in order to get a non-vanishing contribution from
integration around a saddle point $\Phi$ it must be invariant under
the group $H_1$ generated by $Q_1$ and $\wh L_0$.
Furthermore after taking into account appropriate
measure factors we can express the 
path integral as integration over
an $H_1$ invariant slice passing through $\Phi$.

One might wonder whether it is possible to argue that the
path integral can be simultaneously localized into the subspace that
is invariant under all the $Q_i$'s \i.e.\ the subgroup $H_0$.
An intuitive argument to this effect can be given as follows.
We have chosen to define the path integral by first integrating over
the orbits of $H_1$, then integrating over the orbits of the rest
of the elements of $H_0$ and finally integrating over the rest of
the variables in some order. Now since the bosonic subgroup of
$H_0$ -- generated by $(\wh L_0- J^3)$ -- is compact, integration
along the orbit of this generator cannot produce a divergence.
On the other hand integration over the orbit of any of the fermionic
generators $Q_i$ will produce a zero. Thus unless the configuration is
invariant under all the fermionic generators $Q_i$ of $H_0$, 
the contribution
to the path integral from the orbits of $H_0$ through this configuration
will vanish. 

One however runs into problem in trying to construct a formal 
proof of this intuitive expectation.
Naively one could proceed by first showing localization 
under $Q_1$ as we have described above and then adding
further terms to the action to  localize  the integral into
subspaces invariant under the other $Q_i$'s. 
The problem with this approach is that generically the extra insertion
$\exp(-t Q_1 F) $ into the path integral in \refb{esk4} may not
maintain
invariance under the other $Q_i$'s. 
As a result the $Q_i$'s are not
symmetries of the extra factor given in \refb{esk9}.
Similar problem  arises for the factor \refb{eaddz}.
We believe that this is essentially a technical issue 
and the path integral
really receives contribution only from the $H_0$ invariant 
configurations. However we shall take a conservative approach
and use only the requirement of $H_1$ invariance in our
subsequent analysis.
In particular the analysis of \S\ref{sorbit}, showing that the
infinite set of fermionic and bosonic zero modes arising from the
asymptotic symmetries of $AdS_2$ give a finite contribution to the
path integral, will rely only on localization onto $(\wh L_0-J^3)$
invariant subspace. In \S\ref{sexam} we shall describe
freely acting orbifolds of the original $AdS_2\times S^2$
background which contribute to the path integral as
new saddle points. Localization requires us to look for
$H_1$ invariant orbifolds. 
However as we shall see, the requirement
of $H_1$ invariance automatically forces us to have orbifolds
invariant under the full $H_0$ subgroup of $SU(1,1|2)$.

\sectiono{Integrating Over the
Orbit of the  Superconformal Current Algebra} \label{sorbit}

String theory on $AdS_2\times S^2$ space, 
describing the near horizon geometry of a BPS black hole,
has an infinite group of asymptotic symmetries besides the
global $SU(1,1|2)$ transformations which leave the
$AdS_2\times S^2$ background invariant. These more general
transformations do not leave the $AdS_2\times S^2$ background
invariant but preserve the asymptotic condition on the various
fields. Hence they can be used to generate new solutions from
a given solution. As was shown in \cite{0903.1477}, the Euclidean
action of the theory remains unchanged under these transformations
even after taking into account the effect of the infrared cut-off.
Thus they represent zero modes. In a non-supersymmetric theory
where only bosonic zero modes are present,
integration over these zero modes
will generate an infinite factor in the partition function. Hence
integration over these directions must be restricted by declaring
the corresponding transformations 
as gauge transformations. However as was pointed out
in \cite{0903.1477}, in a supersymmetric 
theory there is a possibility of
cancellation between the bosonic and fermionic zero mode
integrals yielding a finite result. We shall now demonstrate
that this is indeed what happens.

The generators of the extended superconfomal algebra may be
labelled as $\wt L_n$, $\wt J^i_n$ and $\wt G^{\alpha\beta}_r$
with $n\in \ZZZ$, $r\in \ZZZ+{1\over 2}$, $1\le i\le 3$ and
$\alpha, \beta=\pm$. The generators of   $su(1,1|2)$ discussed
in \S\ref{ssymm} are special cases of these generators with the
identification
\be \label{eid1}
\wh L_0 =\wt L_0, \quad
\wh L_\pm = \wt L_{\mp 1}, \quad J^i = \wt J^i_0,
\quad \wh G^{\alpha\beta}_\pm = 
\wt G^{\alpha\beta}_{\mp{1\over 2}}\, .
\ee
For our analysis we shall not need the full superconformal current
algebra, but only the commutators of the various generators
with $\wh L_0$ and $J^3$. They are given by
\ben \label{eid2}
&& [\wh L_0, \wt L_n] = -n \, \wt L_n, \quad [\wh L_0, \wt J^i_n]
= -n \, \wt J^i_n, \quad [\wh L_0, \wt G^{\alpha\beta}_r]
= -r \, \wt G^{\alpha\beta}_r, \nonumber \\
&& [J^3, \wt L_n] =[J^3, \wt J^3_n]= 0, \quad 
[J^3, \wt J^\pm_{n}]=\pm \wt J^\pm_{n}, \quad
[J^3, \wt G^{\alpha\beta}_r] = {1\over 2} \, \beta\, 
\wt G^{\alpha\beta}_r, \nn &&
\qquad \qquad \wt J^\pm_n\equiv \wt J^1_n\pm
i \wt J^2_n\, .
\een
This gives
\ben \label{eid3}
&& [\wh L_0 - J^3, \wt L_n] = -n \, \wt L_n, \quad
[\wh L_0 - J^3, \wt J^3_n]=-n\, \wt J^3_n, \quad [\wh L_0 - J^3, 
\wt J^\pm_{n}]= (-n \mp 1) \, \wt J^\pm_{n}, \nn &&
[\wh L_0 - J^3, \wt G^{\alpha\beta}_r]=
\left(-{1\over 2}\beta - r\right) \wt G^{\alpha\beta}_r\, .
\een

Consider now an $H_1$-invariant saddle point
and analyze the contribution from the zero modes generated by the
action of the
superconformal algebra. First note that most of the modes
generated by the superconforml algebra carry non-zero eigenvalues
under $\wh L_0-J^3$. They are part of the deformations labelled
by $z^s_{(m)}$ and $\zeta^s_{(m)}$ in \S\ref{slocal} 
and are eliminated
by the localization procedure described in \S\ref{slocal}.
Thus we only need to worry about deformations generated by
$\wh L_0 - J^3$ invariant generators. Of these several are part
of the global symmetry group $SU(1,1|2)$ and have already been
taken into account in the analysis of \S\ref{slocal}.
{}From \refb{eid3} we see that the only $\wh L_0-J^3$
invariant generators which are not part of
$SU(1,1|2)$ are 
$\wt J^+_{-1}$ and $\wt J^-_{1}$. Since together with 
$\wt J^3_0=J^3$
they generate an $SU(2)$ group, the integration over these
zero modes will give us a finite factor proportional to the
volume of $SU(2)$. This shows that around an $H_1$-invariant
saddle point,
integration over the fermionic and bosonic zero modes generated
by the full superconformal current algebra gives a finite result. 

\sectiono{Examples of $H_1$-invariant Saddle Points}
\label{sexam}

In this section we shall review the construction of a class of saddle
points from orbifolds of the near horizon geometry of the black
hole\cite{0810.3472,0903.1477,0904.4253} and verify their
$H_1$-invariance.
We shall focus on type IIB
string theory on $K3$ -- the theory discussed in
appendix \ref{killingspinor}
-- and consider six dimensional geometries 
whose asymptotic form coincide
with that of  $S^1\times \wt S^1\times
AdS_2\times S^2$ with background 3-form 
fluxes.\footnote{Note
$S^1$ and $\wt S^1$ are not factored metrically, \i.e.\ we allow
the metric to have components 
which mix $S^1$ and $\wt S^1$ coordinates.}
The simplest $H_1$-invariant saddle point is
$S^1\times \wt S^1\times
AdS_2\times S^2$ with background fluxes, given in \refb{eb1}:
\ben \label{eb1pre}
ds^2 &=& v \, \left( d\eta^2 +\sinh^2\eta  d\theta^2\right)
+ u (d\psi^2 + \sin^2\psi d\phi^2) + {R^2\over\tau_2} 
\left|dx^4+\tau dx^5
\right|^2 \, , 
\nn 
G^I &=& {1\over 8\pi^2}\left[
Q_I  \, \sin\psi\, dx^5\wedge d\psi\wedge d\phi +
P_I \, \sin\psi\, dx^4\wedge d\psi\wedge d\phi + \hbox{dual}
\right]\, , \nn
V_I^{~i}&=&\hbox{constant}, \quad V_I^{~r}=\hbox{constant}\, .
\een
As discussed in 
appendix
\ref{killingspinor}, this background is invariant under the full
$SU(1,1|2)$ symmetry group. The classical contribution to the
quantum entropy function from this saddle point is given by
$\exp(S_{wald})$ where $S_{wald}$ denotes the classical
Wald entropy\cite{9307038}.

We shall now construct other $H_1$ invariant saddle points 
with the same asymptotic behaviour as \refb{eb1pre}
by
taking orbifold of the above background by some discrete $\ZZZ_s$
group. Since $H_1$ is generated by $Q_1$ and $Q_1^2$,
in order to preserve $H_1$ the $\ZZZ_s$ 
action must commute
with $Q_1$. Typically the generator
of the $\ZZZ_s$ transformation will involve
an element of $SU(1,1|2)$ together with an internal symmetry
transformation that commutes with $SU(1,1|2)$.
Now one can see from the algebra \refb{ekx7} that the
only bosonic generator of $su(1,1|2)$ that commutes with $Q_1$
is $(\wh L_0-J^3)$. Thus the part of the orbifold group generator
that belongs to $SU(1,1|2)$ must be an element of 
the $U(1)$ subgroup
generated by $(\wh L_0-J^3)$. However
since $(\wh L_0-J^3)$ commutes
with the $H_0\times U(1)$ subgroup of $SU(1,1|2)$ generated by
$Q_1,\cdots Q_4$ and $(\wh L_0\pm J^3)$, we see that any such
saddle point will automatically also
be invariant under this bigger subgroup of $SU(1,1|2)$.
We shall now give some specific examples of such orbifolds.

It was shown in 
\cite{0712.0043,0801.0149} that
with the help of a duality transformation
we can bring the charge vector to the form
\be \label{edefl}
(Q,P)=(\ell Q_0, P_0)\, ,
\ee 
for some integer $\ell$, representing a 
duality invariant combination 
of the charges\cite{0702150}.
Here $(Q_0,P_0)$ are primitive vectors
of the charge lattice,
satisfying
\be \label{eb3}
\gcd\left(\{Q_{0I}P_{0J}-Q_{0J}P_{0I}\}\right) = 1\, .
\ee
We now consider an orbifold of the background \refb{eb1pre} by
the $\ZZZ_s$ transformation\cite{0903.1477}
\be \label{ech4}
(\theta,\phi,x^5)\to \left(\theta+{2\pi\over s},\phi-{2\pi 
\over s},x^5
+{2\pi k\over s}
\right)\, , \quad k,s\in \ZZZ, \quad \gcd(s,k)=1\, .
\ee
Since the circle parametrized by 
$x^5$ is non-contractible, this is a
freely acting orbifold. At the origin $\eta=0$ of the $AdS_2$ space
we have a non-contractible 3-cycle spanned by $(x^5,\psi,\phi)$,
with the identification 
$(x^5,\psi,\phi)=(x^5+2\pi k/s, \psi, \phi-2\pi/s)$.
As a result of this identification the total flux of $G^I$ through
this cycle is equal to $Q_I/s=(l/s)Q_{0I}$. 
Since the flux quantization
constraints require the fluxes through this new 3-cycle to be integers,
we see that this orbifold is an allowed configuration in string theory
only when $\ell/s$ is an integer.

Since  $(\wh L_0-J^3)$ shifts $\theta$ and $\phi$ in opposite directions, 
the $\ZZZ_s$ transformation described in \refb{ech4} is
generated by $(\wh L_0-J^3)$ together with a shift along
$x^5$. Since all the generators of $SU(1,1|2)$ are invariant under a
shift along $x^5$, we see that the subgroup of $SU(1,1|2)$ that commutes
with $(\wh L_0-J^3)$ will be a symmetry of this orbifold. This
is precisely the group $H_0$ together with the
$U(1)$ subgroup generated by $(\wh L_0+J^3)$. 
Indeed one can verify explicitly
that the Killing spinors associated with the generators
$G^{++}_+$, $G^{--}_-$, $G^{-+}_+$ and $G^{+-}_-$
described in \refb{ekillsoln}
are invariant under the transformation
\refb{ech4}.

It was shown in 
\cite{0903.1477} that the 
orbifold described above has the correct asymptotic behaviour.
For this we rename the coordinates 
$(\eta,\theta,\phi,x^5)$ appearing in
\refb{eb1} as $(\wt\eta,\wt\theta,\wt\phi,\wt x^5)$
and express the new configuration as
\ben \label{enewcon}
ds^2 &=& v \, \left( d\wt\eta^2 +\sinh^2\wt\eta  d\wt\theta^2\right)
+ u (d\psi^2 + \sin^2\psi d\wt\phi^2) + {R^2\over\tau_2} 
\left|dx^4+\tau d\wt x^5
\right|^2 \, , \nn
G^I &=& {1\over 8\pi^2}\left[
Q_I  \, \sin\psi\, d\wt x^5\wedge d\psi\wedge d\wt \phi +
P_I \, \sin\psi\, dx^4\wedge d\psi\wedge d\wt \phi + \hbox{dual}
\right]\, , \nn
(\wt\theta,\wt\phi,\wt x^5) &\equiv& \left(\wt\theta+{2\pi\over s},
\wt\phi-{2\pi 
\over s},\wt x^5
+{2\pi k\over s}
\right)
\equiv (\wt\theta+2\pi,\wt\phi,\wt x^5)\nn
&\equiv& (\wt\theta,\wt\phi+2\pi,\wt x^5)
\equiv (\wt\theta,\wt\phi,\wt x^5+2\pi)\, .
\een
We now make the coordinate transformation:
\be \label{ectrs}
\theta = s\wt\theta, \quad \phi = \wt\phi+(1-s)\wt\theta , 
\quad x^5 = \wt x^5 - k\wt\theta, \quad \eta=\wt\eta -\ln s\, .
\ee
In these coordinates the background \refb{enewcon}
takes the form
\ben \label{emodcon}
ds^2 &=& v \, \left( d \eta^2 +\sinh^2 \eta  
\left( 1 + {(1-s^{-2})e^{-\eta}\over 2\sinh\eta}\right)^2\,
d \theta^2\right)
+ u (d\psi^2 + \sin^2\psi (d\phi+d\theta-s^{-1}d\theta)^2)\nn
&& + {R^2\over\tau_2} 
\left|dx^4+\tau (dx^5 + k s^{-1}d\theta)
\right|^2 \, , \nn
G^I &=& {1\over 8\pi^2}\left[
Q_I  \, \sin\psi\, (dx^5 + k s^{-1}d\theta)
\wedge d\psi\wedge (d\phi+d\theta-s^{-1}d\theta)\right. 
\nn && \left. +
P_I \, \sin\psi\, dx^4\wedge d\psi\wedge 
(d\phi+d\theta-s^{-1}d\theta) 
+ \hbox{dual}
\right]\, , \nn
&& (\theta,\phi,x^5)\equiv \left(\theta+{2\pi},
\phi,x^5
\right)\equiv \left(\theta,
\phi+{2\pi},x^5
\right)\equiv \left(\theta,
\phi,x^5+{2\pi}
\right)\, .
\een
Since the asymptotic region lies at large $\eta$, we see that
this has the same asymptotic behaviour as the $S^1\times \wt S^1
\times AdS_2\times S^2$ background described in 
\refb{eb1}. Note the presence
of the $d\theta-s^{-1}d\theta$ terms added to $d\phi$ and
$k s^{-1}d\theta$ terms added to $dx^5$. From the point of
view of the two dimensional theory living on $AdS_2$ these
represent constant values of the gauge fields arising from
the $5\theta$ and $\phi\theta$ components of the metric.
As discussed in detail in 
\cite{0809.3304,0810.3472,0903.1477}, 
in defining the path integral
over $AdS_2$ we must integrate over these modes.
Thus \refb{emodcon} is an allowed configuration over
which the path integral should be performed.
The classical 
contribution to the quantum entropy
function from this saddle point is given by
$\exp(S_{wald}/s)$\cite{0903.1477}. 
These match with the asymptotic behaviour of
specific extra terms in the microscopic formula which appear
when the integer $\ell$ introduced in \refb{edefl} is larger than
1.

Starting with the Killing spinors given in \refb{ekillsoln}
with $(\theta,\phi,\eta)$ replaced by $(\wt\theta,\wt\phi,\wt\eta)$,
and then using the coordinate transformations given in 
\refb{ectrs} one can verify that the Killing spinors corresponding
to $G^{++}_+$, $G^{--}_- $, $G^{-+}_+$ and $G^{+-}_-$ are
given by the same expressions as in \refb{ekillsoln} with
$\eta$ replaced by $\eta +\ln s$. {}From the structure of
\refb{ekillsoln} it can be seen that for large $\eta$ the
replacement of $\eta$ by $\eta +\ln s$ multiplies the Killing
spinors by an overall factor of $\sqrt s$. This is just an overall
normalization constant and can be removed. Thus we see that
asymptotically the Killing spinors of this new saddle point
coincide with those of the background \refb{eb1pre}.
The regularity of the Killing spinors at the origin follows from
the fact that the new saddle point is obtained as a freely
acting orbifold of \refb{eb1pre} and that in the parent theory
the Killing spinors were regular everywhere.

Finally we can consider another class of orbifolds for which
$k$ appearing in \refb{ech4} vanishes, or more generally,
has a common factor with $s$. 
The orbifold group still commutes with $H_0$ and hence we
expect $H_0$ to be a symmetry of this orbifold. However
in this case the orbifold action has fixed points and we no longer
have a freely acting orbifold.
Let us consider the $k=0$ case for 
definiteness\cite{0810.3472}. The points
$(\eta=0; \psi=0,\pi)$ are fixed points of this orbifold group, and
the 3-cycles spanned by $(x^5,\psi,\phi)$ and
$(x^4,\psi,\phi)$ at $\eta=0$ now pass through these fixed points.
The fluxes through these three cycles from regions outside the
fixed points are given by $Q_I/s$ and $P_I/s$ respectively. However
 flux quantization rule does not put any constraints on the charge
vectors $Q_I$ and $P_I$. Instead it requires that there must be
additional flux at the fixed points which make the total
flux through these 3-cycles satisfy the correct quantization 
rules.\footnote{Such fluxes have been considered before in
\cite{9507027} in a different context.} As was argued in
\cite{0810.3472}, the contribution to the partition function from
these saddle points is given by $\exp(S_{wald}/s)$ if we ignore
the contribution from the fixed points. Furthermore 
the contribution from the fixed points add at most
constants of order unity to $S_{wald}/s$ whereas $S_{wald}$
grows quadratically with the charges carried by the black hole.
Thus for large charges the contribution from the fixed points
to the exponent is subleading. 

In a dual description of these theories in $M$-theory the near
horizon geometry of these black holes can have an extra circle
that combines with the $AdS_2$ to give a locally $AdS_3$ space.
In this case one can get freely acting $\ZZZ_s$
orbifolds by accompanying
the orbifold action by a translation along this extra 
circle\cite{0904.4253}, without
imposing any additional arithmetic condition on the 
charges of the type $\ell/s\in\ZZZ$. 
This could provide a possible way to
analyze the orbifolds with fixed points in the type IIB description.

\sectiono{Comments} \label{scomm}

In this paper we have used the localization procedure to classify
the saddle points which will contribute to string theory 
path integral over the near horizon geometry of extremal BPS
black holes. This path integral is required for the computation of
quantum entropy function, which appears in the macroscopic
computation of the entropy of extremal black holes via 
$AdS_2/CFT_1$ correspondence\cite{0809.3304}.

We hope that the same localization techniques will also simplify the
computation of the path integral around each of the saddle points,
{\it e.g.} by reducing the path integral over the fields to a finite
dimensional integral.
In particular for quarter BPS black holes in type IIB string theory
on $K3\times T^2$ if the contribution to the path integral from some
of the saddle points can be expressed as finite dimensional
integrals, they can then be compared
with the corresponding microscopic results derived in
\cite{9607026,0412287,0505094,0506249,0605210,
0802.0544,0802.1556,0803.2692}, providing us with a precision
test of the $AdS_2/CFT_1$ correspondence. The formulation of
string theory on $AdS_2\times S^2$ described in
\cite{9907200} could also be a useful tool
in this venture.

Finally we note that drawing inspiration from
\cite{0501015} a recent paper\cite{0904.4486} expressed the
expectation value of circular 't Hooft - Wilson  
loop operators in an $\NN=4$
supersymmetric gauge theory as a path integral
over the field theory on $AdS_2\times S^2$ background. Except for
the replacement of the string theory by $\NN=4$ supersymmetric
Yang-Mills theory, this path integral is identical to what appears
in the definition of the quantum entropy function. Thus we expect
that any method (like the one in the present paper)
developed for the study of quantum entropy function is likely
to be useful for the study of the 't Hooft - Wilson
loop operators in
$\NN=4$ super-Yang-Mills theory. Similarly
any method developed for computing 't Hooft - Wilson
loop operators in
$\NN=4$ super-Yang Mills theory (like the one developed in
\cite{0904.4486}) may be useful for the 
computation of quantum entropy
function in string theory. It will also be useful to explore whether
the correspondence between the 't Hooft - Wilson
loop and the quantum
entropy function is just a mathematical coincidence or whether
there is some deeper physical reason behind it.

\bigskip

{\bf Acknowledgement:} We would like to thank 
Sayantani Bhattacharyya, Chethan
Gowdigere, Dileep Jatkar and Yogesh Srivastava 
for useful discussions.

\appendix

\sectiono{Killing Spinors in Six Dimensional Supergravity
on $S^1\times \wt S^1\times AdS_2\times S^2$} 
\label{killingspinor}

In this appendix we shall analyze the Killing spinors in six
dimensional $\NN=4$ chiral supergravity compactified
on $S^1\times \wt S^1\times AdS_2\times S^2$. This theory
is dual to M-theory on $K3\times T^3\times AdS_2\times S^2$,
for which the Killing spinor equations have been analyzed in
\cite{0608021}. Thus we
could try to recover our answer by dualizing the results of
\cite{0608021}. We shall however analyze 
the Killing spinor equations
directly in the six dimensional chiral supergravity in the presence
of arbitrary background fluxes. This will make the duality covariance
of the equations manifest.

We begin with the six dimensional supergravity theory obtained
by dimensional reduction of type IIB supergravity on
$K3$\cite{romans,9712176}. We shall follow the conventions
of \cite{9804166}.
The bosonic fields in the theory are the metric $g_{MN}$,
matrix valued
scalar fields $V_I^{~i}$, $V_I^{~r}$ ($1\le i\le 5$,
$6\le r\le 26$) satisfying
\be \label{evcond}
V L V^T = L, \qquad L=diag(+^5, -^{21})\, ,
\ee
and 2-form fields
$B^I_{MN}$ ($1\le I\le 26$) with field strengths
$G^I=dB^I$ satisfying the following self duality
constraint:
\be \label{eself}
H^i_{MNP} = {1\over 3!} \, e_{MNPQRS}\, H^{iQRS},
\qquad H^r_{MNP} = -{1\over 3!} \, e_{MNPQRS}\, H^{rQRS}
\, ,
\ee
where
\be \label{edefh}
H^i_{MNP} = G^I_{MNP} V_I^{~i}, \qquad 
H^r_{MNP} = G^I_{MNP} V_I^{~r}\, .
\ee
$e_{MNPQRS}$ is a six form defined via
\be \label{edefe}
e^{MNPQRS} =  |\det g|^{-1/2} \, \eps^{MNPQRS}\, ,
\ee
$\eps$ being the totally antisymmetric symbol. 
We shall label the  time coordinate by
$t$ and the space-coordinates by $(x^4,x^5,\eta,\psi,\phi)$
and choose the convention
\be \label{eeconv}
\eps^{t45\eta\psi\phi} = 1\, .
\ee
Indices of $e$ are raised and lowered by the metric $g_{MN}$.
Not all components of $V$ describe physical degrees
of freedom since there is an identification
\be \label{eidv}
V\equiv V \,O\, ,
\ee where
$O$ is an $SO(5)\times SO(21)$ matrix acting on the first five and
the last twenty one indices respectively.

In the sector where the bosonic fields are taken to be space-time
independent constants, the equations of motion take the form
\ben \label{eeom}
& R_{MN} = H^i_{MPQ} \, H_N^{iPQ} + 
H^r_{MPQ} \, H_N^{rPQ}\, \nn
& H^i_{MNP} H^{rMNP} = 0\, ,
\een
where $R_{MN}$ is the Ricci tensor defined in the 
sign convention
in which on the sphere the Ricci scalar $g^{MN}R_{MN}$
is positive.
We now look for a solution in this theory of the form
\ben \label{eb1lor}
ds^2 &=& v \, \left( d\eta^2 -\sinh^2\eta  dt^2\right)
+ u \, \left(d\psi^2 + \sin^2\psi d\phi^2\right) + {R^2\over\tau_2} 
\left|dx^4+\tau dx^5
\right|^2 \, , \nn
G^I &\equiv& {1\over 3!} G^I_{MNP} dx^M\wedge dx^N\wedge
dx^P\nn
&=& {1\over 8\pi^2}\left[
Q_I  \, \sin\psi\, dx^5\wedge d\psi\wedge d\phi +
P_I \, \sin\psi\, dx^4\wedge d\psi\wedge d\phi + \hbox{dual}
\right]\, , \nn
 V_I^{~i}&=&\hbox{constant}, \quad V_I^{~r}=\hbox{constant}\, .
 \een
Here `dual' denotes the dual 3-form required to make $G^I$
satisfy the self-duality constraint given in \refb{eself},
$v$, $u$, $R$  are real constants
and $\tau=\tau_1+i\tau_2$ is a complex constant. 
$(\eta,t)$ label an $AdS_2$ space,
$(\psi,\phi)$ label a 2-sphere and $x^4,x^5$ label coordinates
along $\wt S^1$ and $S^1$ respectively, each taken to have
period $2\pi$.
$Q_I$ and $P_I$ denote the fluxes through the 3-cycles
$S^1\times S^2$ and $\wt S^1\times S^2$ respectively,
and are related to the integer charges carried by the
black hole whose near horizon geometry is described by
\refb{eb1lor}.
In order to solve \refb{eeom} we note that given any charge
vectors $(Q,P)$ satisfying
\be \label{eq2p2}
Q^2 > 0, \quad P^2 >0, \quad Q^2 P^2 > (Q\cdot P)^2\, ,
\ee
where 
\be \label{edefq2}
Q^2 =  Q^T L Q, \quad P^2=P^T L P, \quad Q\cdot P
= Q^T L P\, ,
\ee
we can always find a matrix $S$ 
satisfying $SLS^T=L$ such that
\be \label{eqpc}
Q=S Q_0, \quad P = S P_0\, ,
\ee
where
\be \label{eq0p0}
Q_0 = \pmatrix{Q\cdot P/\sqrt{P^2}\cr \cr
\sqrt{Q^2 P^2 - (Q\cdot P)^2}/\sqrt{P^2}\cr \cr
0\cr\cdot\cr \cdot}
\, , \qquad P_0 = \pmatrix{\sqrt{P^2}\cr\cr
0\cr\cr 0\cr\cdot\cr \cdot}\, .
\ee
In that case eqs.\refb{eeom} is solved by \refb{eb1lor} for the
choice
\ben \label{esoln}
&& V = (S^T)^{-1}, \quad 
\tau_1=Q\cdot P/ P^2, \quad
\tau_2 = \sqrt{Q^2 P^2 - (Q\cdot P)^2}/P^2, \nn
&& v = u = {1\over 16\pi^4 R^2}\, 
\sqrt{Q^2 P^2 - (Q\cdot P)^2}.
\een
Using eq.\refb{edefh} this gives
\be \label{ehres}
H^i = {1\over 8\pi^2}\left[
Q_0^i  \, \sin\psi\, dx^5\wedge d\psi\wedge d\phi +
P_0^i \, \sin\psi\, dx^4\wedge d\psi\wedge d\phi + \hbox{dual}
\right], \quad H^r = 0\, .
\ee
Note that $R$ is arbitrary. Furthermore $S$ defined through
\refb{eqpc} is ambiguous up to an $SO(3,21)$ transformation
from the right  acting on the last 24 elements. Thus 
$V$ given in \refb{esoln} is  determined only up to an
$SO(3,21)$ multiplication from the right. Due to the
identification \refb{eidv}
only an
$SO(3,21)/SO(3)\times SO(21)$ family of these describe
physically inequivalent configurations. These parameters which
are left undetermined by the equations of motion describe
flat directions of the entropy function.

The fermion fields in this theory consist of a set of
gravitini $\psi_M$ and a set of spin 1/2 fermions $\chi^r$.
$\chi^r$ transforms as {\bf 21}  of $SO(21)$, {\bf 4} of 
$SO(5)$ and a right chiral spinor of $SO(5,1)$
where $SO(5,1)$ denotes the
tangent space Lorentz group, $SO(21)$ is the
internal symmetry
group acting on the index $r$,
 and $SO(5)$ is the internal symmetry
group acting on the index $i$. 
In what follows we shall suppress all the $SO(5)\times SO(5,1)$
spinor indices.
For each $M$, $\psi_M$ 
transforms as {\bf 4}
of $SO(5)$ and a left-chiral spinor of $SO(5,1)$. Finally
the supersymmetry transformation parameter $\eps$ 
transforms as
a {\bf 4}  of $SO(5)$ and a left chiral spinor
of $SO(5,1)$.
Let us denote
the vielbeins by $e_M^{~A}$ with $A$ labelling an
$SO(5,1)$ tangent space index, the SO(5,1) gamma matrices
by $\wt\Gamma^A$ and the SO(5) gamma matrices by
$\wh \Gamma^i$. We shall also use the symbol $\Gamma^M$
to denote the $SO(5,1)$ gamma matrices in the coordinate
basis, \i.e.\ we have
\be \label{ecorb}
\wt\Gamma^A = e_M^{~A}\, \Gamma^M\, .
\ee
Then the $SO(5,1)$ chirality conditions on various spinors
may be described as
\ben \label{echiralcond}
&&\left(
\Gamma^{QRS} -{1\over 3!} e^{MNPQRS} \Gamma_{MNP}
\right) \chi^r=0,\nn
&&\left(
\Gamma^{QRS} +{1\over 3!} e^{MNPQRS} \Gamma_{MNP}
\right) \psi_K=0, \nn
&&\left(
\Gamma^{QRS} +{1\over 3!} e^{MNPQRS} \Gamma_{MNP}
\right) \eps =0,
\een
where 
\be \label{egsymm}
\Gamma^{M_1\cdots M_k}={1\over k!} \left(\Gamma^{M_1}
\cdots \Gamma^{M_k} + \hbox{permutations with sign}
\right)\, .
\ee
Besides this all the spinors $\psi_M$, $\chi^r$ and $\eps$ satisfy
the symplectic Majorana condition, {\it e.g.} we have
\be \label{esymplectic}
\bar \eps = \eps^T \, C\, \Omega\, ,
\ee
where $\Omega$ is the SO(5) charge conjugation matrix
acting on the $SO(5)$ spinor index and 
$C$ is the SO(5,1) charge conjugation matrix 
acting on the SO(5,1)
spinor index.
The supersymmetry transformation
laws of various fields take the form
\ben \label{esusy}
&& \delta e_M^{~A} =\bar\eps \wt\Gamma^A \psi_M \nn
&& \delta \psi_M = D_M\epsilon -{1\over 4}
H^i_{MNP} \Gamma^{NP} \wh\Gamma^i \eps, 
\qquad D_M\eps \equiv \p_M \eps +{1\over 4} \omega_M^{AB}
\wt\Gamma_{AB}\eps -{1\over 4}\, Q_M^{ij}\, \wh \Gamma^{ij}
\eps\, ,\nn
&& \quad \omega_M^{AB} \equiv -g^{NP} e_N^{~B} \p_M 
e_P^{~A} +  e_N^{~A} e_P^{~B} g^{PQ} \Gamma^N_{QM},
\qquad \Gamma^M_{NP} \equiv {1\over 2}\, g^{MR}\,
(\p_N g_{PR} + \p_P g_{NR} -\p_R g_{NP})\, , \nn
&& \delta B^I_{MN} = - V^{Ii} \bar\eps \Gamma_{[M}
\wh\Gamma^i\psi_{N]} +{1\over 2} V^{Ir}
\bar\eps \Gamma_{MN} \chi^r, \nn
&& \delta \chi^r = {1\over \sqrt 2} \Gamma^M P_M^{~ir}
\wh\Gamma^i\eps +{1\over 12} \Gamma^{MNP} H^r_{MNP}
\eps, \nn
&& \delta V_I^{~i} = \bar\eps \wh\Gamma^i \chi^r V_I^{~r}\, ,
\nn
&& \delta V_I^{~r} =\bar\eps \, \wh\Gamma^i \chi^r V_I^{~i}\, , 
\een
where the index $I$ is raised and lowered by the matrix $L$ and
\be \label{epdef}
P_M^{~ir} = {1\over \sqrt 2}\, \p_M V_I^{~i} (V^{-1})_r^{~I}\, ,
\qquad 
Q_M^{~ij} = \p_M V_I^{~i} (V^{-1})_j^{~I}\, .
\ee
Thus the Killing spinor equations, obtained by setting the
variation of $\chi^r$ and $\psi_M$ to zero, are given by
\ben \label{ekillfull}
&& D_M\epsilon -{1\over 4}
H^i_{MNP} \Gamma^{NP} \wh\Gamma^i \eps=0, \nn
&& 
{1\over \sqrt 2} \Gamma^M P_M^{~ir}
\wh\Gamma^i\eps +{1\over 12} \Gamma^{MNP} H^r_{MNP}
\eps=0\, .
\een

We shall try to solve these equations in the background
\refb{eb1lor}, \refb{esoln}. The analysis simplifies if we
note that in this background
\be \label{eback}
P_M^{~ir}=0, \qquad Q_M^{ij}=0\, , \qquad H^r_{MNP}=0\, .
\ee
Thus the second set of equations in \refb{ekillfull} are
satisfied automatically. 
The first set of equations can be split
into two sets by taking $M=(4,5)$ and $M=(\eta, t, \psi, \phi)$:
\ben \label{esets}
&& H^i_{a\mu\nu} \Gamma^{\mu\nu} \wh
\Gamma^i \eps=0\, ,
\quad a=4,5, \quad \mu,\nu =\eta, t, \psi, \phi\, , \nn
&& D_\mu\eps +{1\over 2} H^i_{a\mu\nu} \Gamma^{a\nu}
\wh\Gamma^i\eps=0\, .
\een

Since we shall eventually be interested in finding the Killing
spinors in the euclidean theory, we shall
now make a euclidean continuation of the theory. This is
done
by making the replacement
\be \label{eeuc}
t\to -i\theta\, ,
\ee
and replacing \refb{edefe}, \refb{eeconv}
by
\be \label{edefenew}
e^{MNPQRS} = i|\det g|^{-1/2} \, \eps^{MNPQRS}\, ,
\qquad \eps^{\theta 45\eta\psi\phi} = 1\, .
\ee
This will guarantee that a solution obtained by euclidean
rotation of a Minkowski solution will satisfy the self-duality
conditions \refb{eself} with $e_{MNPQRS}$ defined via
\refb{edefenew}.
Furthermore the chirality projection rules \refb{echiralcond},
the supersymmetry transformation rules \refb{esusy}
and the killing spinor equations \refb{ekillfull} all remain
unchanged as long as we use the new definition
\refb{edefenew}. Finally since the {\bf 4} representation of
$SO(6)$ is different from its conjugate representation
${\bf \bar 4}$, we can no longer impose the symplectic Majorana
condition on the spinors. However we shall now take \refb{esymplectic}
as the definition of ${\bar \eps}$ appearing in the supersymmetry
transformation laws. Equivalently, we could first
replace $\bar \eps$ in the supersymmetry transformation laws in
terms of $\eps$ using \refb{esymplectic}, and then
make the Euclidean continuation.
The charge conjugation matrices $C$ and $\Omega$ have to be chosen
so that $\eps_1^T \, C\, \Omega \, \wt\Gamma^A \, \eps_2$
and $\eps_1^T \, C\, \Omega \, \wh\Gamma^i \, \eps_2$
transform as $SO(6)$ vectors and $SO(5)$ vectors respectively
for arbitrary $\eps_1$ and $\eps_2$.

Under the euclidean continuation 
the solution given in \refb{eb1lor}
takes the form:
\ben \label{eb1}
ds^2 &=& v \, \left( d\eta^2 +\sinh^2\eta \, d\theta^2\right)
+ u (d\psi^2 + \sin^2\psi \, d\phi^2) + {R^2\over\tau_2} 
\left|dx^4+\tau dx^5
\right|^2 \, , 
\nn 
G^I &=& {1\over 8\pi^2}\left[
Q_I  \, \sin\psi\, dx^5\wedge d\psi\wedge d\phi +
P_I \, \sin\psi\, dx^4\wedge d\psi\wedge d\phi + \hbox{dual}
\right]\, , \nn
V_I^{~i}&=&\hbox{constant}, \quad V_I^{~r}=\hbox{constant}\, ,\nn
H^i &=& {1\over 8\pi^2}\left[
Q_0^i  \, \sin\psi\, dx^5\wedge d\psi\wedge d\phi +
P_0^i \, \sin\psi\, dx^4\wedge d\psi\wedge d\phi + \hbox{dual}
\right], \quad H^r = 0\, ,\nn
\een
with the various parameters determined from \refb{esoln}.
The equations \refb{esets} take the form
\ben \label{enewsets}
&& H^i_{a\mu\nu} \Gamma^{\mu\nu} \wh\Gamma^i \eps=0\, ,
\quad a=4,5, \quad \mu,\nu =\eta, \theta, \psi, \phi\, , \nn
&& D_\mu\eps +{1\over 2} H^i_{a\mu\nu} \Gamma^{a\nu}
\wh\Gamma^i\eps=0\, .
\een

Using the self-duality constraints \refb{eself}, the
chirality constraints \refb{echiralcond}, 
the explicit form of the solutions
given in \refb{esoln}, and \refb{eq0p0},
the first set of equations in \refb{enewsets} takes the simple form
\be \label{esimple}
\wh \Gamma^1\eps = \Gamma_{45} \, (\det g^{(45)})^{-1/2}\,
\wh\Gamma^2\eps\, ,
\ee
where $g^{(45)}$ denotes the metric on $S^1\times \wt S^1$.
We shall now use \refb{esimple} to simplify the second set of
equations in \refb{enewsets}. For this we need to choose the
vielbeins $e_M^{~A}$ consistent with the background
\refb{eb1}. We define $e^A\equiv e_M^{~A} dx^M$ and take
\ben \label{etake}
&& e^0 = \sqrt v \sinh\eta \, d\theta , \quad e^1 = 
\sqrt v\, d\eta, \quad
e^2 = \sqrt u \sin\psi \, d\phi, \quad e^3 = \sqrt u d\psi, \nn
&& e^4 = {R\over \sqrt\tau_2}\, (dx^4 + \tau_1 dx^5), \quad
e^5 = R\, \sqrt{\tau_2} \, dx^5\, .
\een
We also denote by $x^m$ for $m=2,3$
the coordinates $(\phi,\psi)$
along $S^2$ and by $x^\alpha$ for $\alpha=0,1$
the coordinates $(\theta,\eta)$ along $AdS^2$. In that case the
second set of equations in \refb{enewsets} are given by
\ben \label{esecond}
&& D_m \eps -{1\over 2} \,\sqrt u \, \ve^{S^2}_{mn} 
\, \Gamma^n
\, \wt \Gamma^4 \wh \Gamma^1 \eps = 0\, , \nn
&& D_\alpha \eps +{i\over 2} \, \sqrt v \, \ve^{AdS_2}_{\alpha\beta} 
\, \Gamma^\beta
\, \wt \Gamma^4 \wh \Gamma^2 \eps = 0\, ,
\een
where $\wt \Gamma^A$ have been defined in 
\refb{ecorb}, and
\be \label{eepsdef}
\ve^{S^2}_{mn} dx^m \wedge dx^n =\sin\psi \,
d\psi\wedge d\phi
\, , \qquad
\ve^{AdS_2}_{\alpha\beta} dx^\alpha \wedge 
dx^\beta =\sinh\eta \, d\eta\wedge d\theta
\, .
\ee

We can analyze these equations by choosing the following representation
of the gamma matrices:
\ben \label{egammarep}
&& \wt \Gamma^0 = \sigma_1\otimes I \otimes I\otimes I\otimes I,
\qquad 
\wt \Gamma^1 = \sigma_2\otimes I \otimes I\otimes I\otimes I,
\qquad 
\wt \Gamma^2 = \sigma_3\otimes \sigma_1
\otimes I\otimes I\otimes I\nn
&& \wt \Gamma^3 = \sigma_3\otimes \sigma_2
\otimes I\otimes I\otimes I, \qquad
\wt \Gamma^4 = \sigma_3\otimes \sigma_3
\otimes \sigma_1\otimes I\otimes I, \qquad
\wt \Gamma^5 = \sigma_3\otimes \sigma_3
\otimes \sigma_2\otimes I\otimes I\nn
&& \wh \Gamma^1 = I\otimes I \otimes I\otimes \sigma_1
\otimes I,
\qquad 
\wh \Gamma^2 = I\otimes I \otimes I\otimes \sigma_2
\otimes I,
\qquad 
\wh \Gamma^3 = I\otimes I \otimes I\otimes \sigma_3
\otimes \sigma_1\nn
&& \wh \Gamma^4 = I\otimes I \otimes I\otimes \sigma_3
\otimes \sigma_2, \qquad
\wh \Gamma^5 = I\otimes I \otimes I\otimes \sigma_3
\otimes \sigma_3\, ,
\een
where the $\sigma_i$ are Pauli matrices and $I$ is the $2\times 2$
identity matrix. 
In this basis the $SO(6)$ charge conjugation matrix $C$ and the 
$SO(5)$ charge conjugation matrix $\Omega$
have the form:
\be \label{eomega}
C=i\sigma_1\otimes \sigma_2\otimes \sigma_1\otimes I
\otimes I, \quad
\Omega=I\otimes I \otimes I\otimes \sigma_1\otimes \sigma_2, 
\ee
so that $C$ and $\Omega$ satisfy respectively the conditions
for $SO(6)$ and $SO(5)$ invariance\footnote{Note that
\refb{eomcond} does not fix the overall phases of $C$ and $\Omega$.
We have chosen them according to our convenience.}
\be \label{eomcond}
(C\wt \Gamma^A)^T = - C\wt\Gamma^A, \qquad
(\Omega\wh\Gamma^{i})^T = -\Omega\wh\Gamma^{i}\, ,
\ee
for all $A$ and $i$.
We now note that the chirality condition 
\refb{echiralcond} and the Killing spinor condition
\refb{esimple} leads to the constraints:
\be \label{efirst}
(\sigma_3\otimes \sigma_3\otimes \sigma_3\otimes I\otimes I)
\, \eps =\eps,
\qquad (I\otimes I\otimes \sigma_3\otimes \sigma_3\otimes I)
\, \eps
=-\eps\, . 
\ee
Due to these constraints we can parametrize $\eps$ by eight
complex parameters $(\{A_i\}, \{B_i\})$:
\ben \label{egenspin}
\eps &=&  \pmatrix{1\cr 0}\otimes \pmatrix{1\cr 0}
\otimes \pmatrix{1\cr 0} \otimes \pmatrix{0\cr 1} \otimes
\pmatrix{A_1\cr B_1}+   \pmatrix{0\cr 1}\otimes \pmatrix{1\cr 0}
\otimes \pmatrix{0\cr 1} \otimes \pmatrix{1\cr 0} \otimes
\pmatrix{A_2\cr B_2} \nn
&& +  \pmatrix{1\cr 0}\otimes \pmatrix{0\cr 1}
\otimes \pmatrix{0\cr 1} \otimes \pmatrix{1\cr 0} \otimes
\pmatrix{A_3\cr B_3} +  \pmatrix{0\cr 1}\otimes \pmatrix{0\cr 1}
\otimes \pmatrix{1\cr 0} \otimes \pmatrix{0\cr 1} \otimes
\pmatrix{A_4\cr B_4}\, . \nn
\een
Further simplification occurs due to the fact that eqs.\refb{esecond}
do not mix the $A_i$'s with $B_i$'s and in fact remain invariant
under the replacement $A_i\leftrightarrow B_i$. Thus we need to
solve the Killing spinor equations in the four dimensional
subspace parametrized by the $A_i$'s (or $B_i$'s). We get eight
solutions $\zeta^{\alpha\beta}_\gamma$
($\alpha,\beta,\gamma=\pm$). We shall first write down the
solutions for $\zeta^{+\beta}_\gamma$. 
All of these solutions have $B_i=0$ and the $A_i$'s
given by:
\ben\label{ekillsoln}
\zeta^{++}_+:   \pmatrix{A_1\cr A_2\cr A_3\cr A_4}
= 2\, v^{1/4}\, e^{i(\theta+\phi)/2} \,  \pmatrix{\sin{\psi\over 2}
\sinh{\eta\over 2}
\cr -\sin{\psi\over 2}\cosh{\eta\over 2}\cr -\cos{\psi\over 2}
\sinh{\eta\over 2}\cr \cos{\psi\over 2}
\cosh{\eta\over 2}}, \nn
 \zeta^{+-}_+:  \pmatrix{A_1\cr A_2\cr A_3\cr A_4}
= 2\, v^{1/4}\, e^{i(\theta-\phi)/2} \,  \pmatrix{-\cos{\psi\over 2}
\sinh{\eta\over 2}\cr \cos{\psi\over 2}
\cosh{\eta\over 2}\cr -\sin{\psi\over 2}\sinh{\eta\over 2}\cr
\sin{\psi\over 2}\cosh{\eta\over 2}}, \nn
 \zeta^{++}_-:  \pmatrix{A_1\cr A_2\cr A_3\cr A_4}
= 2\, v^{1/4}\, e^{-i(\theta-\phi)/2} \,  \pmatrix{-\sin{\psi\over 2}
\cosh{\eta\over 2}\cr
\sin{\psi\over 2}\sinh{\eta\over 2}
\cr \cos{\psi\over 2}
\cosh{\eta\over 2}\cr -\cos{\psi\over 2}
\sinh{\eta\over 2}}, 
\nn \zeta^{+-}_-:  \pmatrix{A_1\cr A_2\cr A_3\cr A_4}
= 2\, v^{1/4}\, e^{-i(\theta+\phi)/2} \,  \pmatrix{\cos{\psi\over 2}
\cosh{\eta\over 2}\cr -\cos{\psi\over 2}
\sinh{\eta\over 2} \cr \sin{\psi\over 2}
\cosh{\eta\over 2}\cr -\sin{\psi\over 2}\sinh{\eta\over 2}}\, . 
\een
The solutions for
$\zeta^{-\beta}_\gamma$ are obtained by replacing the $A_i$'s
by $B_i$'s and vice versa. The normalization factor $2\, v^{1/4}$ has
been included for convenience.

To check the regularity of the Killing spinors at the origin
$\eta=0$ and / or $\psi=0,\pi$, we need to express the
$AdS_2\times S^2$ metric in the $(z,w)$ coordinates as in
\refb{ekx5} and choose the vielbeins as 
\be \label{enewv}
\hat e^0 = {2\, \sqrt v\over 1 - \bar w w}\, dw_I, \quad
\hat e^1 = {2\, \sqrt v\over 1 - \bar w w}\, dw_R, \quad
\hat e^2 = {2\, \sqrt u\over 1 + \bar z z}\, dz_I, \quad
\hat e^3 = {2\, \sqrt u\over 1 + \bar z z}\, dz_R,
\ee
\be \label{edefzr}
w_R+i w_I \equiv w, \quad z_R + i z_I\equiv z\, .
\ee
Since these vielbeins are regular at $w=0$ and / or $z=0$, the
Killing spinors will be regular at these points if they are free
from any singularity in this frame. 
Now using \refb{etake} we get
\ben \label{erot}
&& \hat e^0 = \cos\theta \, e^0 + \sin\theta \, e^1, \quad
\hat e^1 = -\sin\theta \, e^0 + \cos\theta \, e^1,\nn
&& \hat e^2 = \cos\phi \, e^2 + \sin\phi \, e^3, \quad
\hat e^3 = -\sin\phi \, e^2 + \cos\phi \, e^3\, .
\een
The  $\hat e^A$ are related to $e^A$'s by a rotation by
$\theta$ in the 0-1 plane and a rotation by $\phi$ in the
2-3 plane in the tangent space. Since from \refb{egammarep}
we see that on the spinors rotations in the 0-1 plane and
2-3 plane are generated by ${1\over 2}\sigma_3\otimes I
\otimes I\otimes I\otimes I$ and ${1\over 2}I\otimes
\sigma_3\otimes I
\otimes I\otimes I$ respectively, the rotation \refb{erot} is
represented by the matrix
\be \label{erotrep}
\pmatrix{e^{i\theta/2} & \cr & e^{-i\theta/2}}\otimes
\pmatrix{e^{i\phi/2} & \cr & e^{-i\phi/2}}
\otimes I\otimes I\otimes I\, .
\ee
Applying this on \refb{ekillsoln} and using \refb{egenspin}
we get the Killing spinors in the new frame:
\ben\label{ekillhat}
&& \hat\zeta^{++}_+:   \pmatrix{A_1\cr A_2\cr A_3\cr A_4}
= N
  \,  \pmatrix{zw\cr -z\cr -w\cr 1}, \qquad
\hat \zeta^{+-}_+:  \pmatrix{A_1\cr A_2\cr A_3\cr A_4}
= N
  \,  \pmatrix{-w\cr 1\cr -\bar zw\cr \bar z}, \nn
&& \hat \zeta^{++}_-:  \pmatrix{A_1\cr A_2\cr A_3\cr A_4}
= N
  \,  \pmatrix{-z\cr z\bar w\cr 1\cr -\bar w}, \qquad
 \hat
\zeta^{+-}_-:  \pmatrix{A_1\cr A_2\cr A_3\cr A_4}
= N
  \,  \pmatrix{1\cr -\bar w\cr \bar z\cr -\bar z\bar w},
\nn
&& N \equiv {2\, v^{1/4}\over \sqrt{(1+\bar z z)(1-\bar w w)}}\, .
\een
Similar expressions are obtained for $\zeta^{-\alpha}_\beta$ by
replacing the $A_i$'s by $B_i$'s. Eq.\refb{ekillhat} shows that
all the Killing spinors are regular at $z=0$ and / or $w=0$.

If $\theta^{\gamma}_{\alpha\beta}$ 
denotes a  grassman parameter labelling the supersymmetry
transformations, then the supersymmetry
transformation by the spinor 
parameter $\eps=\theta^{\gamma}_{\alpha\beta}
\zeta^{\alpha\beta}_\gamma$
can be identified as the action of $i\theta^{\gamma}_{\alpha\beta}
\wh G^{\alpha\beta}_\gamma$
on various fields. Using the known supersymmetry transformation
rules for various fields given in 
\refb{esusy} and the definition \refb{esymplectic} of $\bar\eps$ 
one finds
\be \label{ecommutator}
\delta_{\eps_2}\delta_{\eps_1} - \delta_{\eps_1}\delta_{\eps_2}
= \eps_2^T \, C\, \Omega\, \Gamma^M\, \eps_1 \, \p_M\, ,
\ee
up to possible gauge transformations of the type given in
\refb{eidv}. Using this we can  verify that commutator of these
supersymmetry generators with themselves and the other symmetries
follow the $su(1,1|2)$ algebra given in \refb{ekx7}.


\begin{thebibliography}{99}

\bibitem{0809.3304}
  A.~Sen,
  ``Quantum Entropy Function from AdS(2)/CFT(1) Correspondence,''
  arXiv:0809.3304 [hep-th].

\bibitem{0810.3472}
  N.~Banerjee, D.~P.~Jatkar and A.~Sen,
  ``Asymptotic Expansion of the N=4 Dyon Degeneracy,''
  arXiv:0810.3472 [hep-th].

\bibitem{0903.1477}
  A.~Sen,
  ``Arithmetic of Quantum Entropy Function,''
  arXiv:0903.1477 [hep-th].

\bibitem{0606244}
  D.~Astefanesei, K.~Goldstein, R.~P.~Jena, A.~Sen and S.~P.~Trivedi,
  ``Rotating attractors,''
  JHEP {\bf 0610}, 058 (2006)
  [arXiv:hep-th/0606244].


\bibitem{0805.0095}
  A.~Sen,
  ``Entropy Function and $AdS_2/CFT_1$ Correspondence,''
  arXiv:0805.0095v4 [hep-th].

\bibitem{0806.0053}
  R.~K.~Gupta and A.~Sen,
  ``Ads(3)/CFT(2) to Ads(2)/CFT(1),''
  JHEP {\bf 0904}, 034 (2009)
  [arXiv:0806.0053 [hep-th]].

 \bibitem{9307038}
  R.~M.~Wald,
  ``Black hole entropy in the Noether charge,''
  Phys.\ Rev.\ D {\bf 48}, 3427 (1993)
  [arXiv:gr-qc/9307038].

\bibitem{9312023}
  T.~Jacobson, G.~Kang and R.~C.~Myers,
  ``On Black Hole Entropy,''
  Phys.\ Rev.\ D {\bf 49}, 6587 (1994)
  [arXiv:gr-qc/9312023].

\bibitem{9403028}
  V.~Iyer and R.~M.~Wald,
  ``Some properties of Noether 
  charge and a proposal for dynamical black hole
  entropy,''
  Phys.\ Rev.\ D {\bf 50}, 846 (1994)
  [arXiv:gr-qc/9403028].

\bibitem{9502009}
  T.~Jacobson, G.~Kang and R.~C.~Myers,
  ``Black hole entropy in higher curvature gravity,''
  arXiv:gr-qc/9502009.

\bibitem{0506177}
  A.~Sen,
  ``Black hole entropy function and the attractor mechanism in higher
  derivative gravity,''
  JHEP {\bf 0509}, 038 (2005)
  [arXiv:hep-th/0506177].

\bibitem{0708.1270}
  A.~Sen,
  ``Black Hole Entropy Function, 
Attractors and Precision Counting of
  Microstates,''
  arXiv:0708.1270 [hep-th].

\bibitem{heckman}
  J.~J.~Duistermaat and G.~J.~Heckman,
  ``On The Variation In The 
  Cohomology Of The Symplectic Form Of The Reduced
  Phase Space,''
  Invent.\ Math.\  {\bf 69}, 259 (1982).

\bibitem{wittens}
  E.~Witten,
  ``Topological Quantum Field Theory,''
  Commun.\ Math.\ Phys.\  {\bf 117}, 353 (1988).

\bibitem{witten92}
  E.~Witten,
  ``The N Matrix Model And Gauged WZW Models,''
  Nucl.\ Phys.\  B {\bf 371}, 191 (1992).

\bibitem{9112056}
  E.~Witten,
  ``Mirror manifolds and topological field theory,''
  arXiv:hep-th/9112056.

\bibitem{9204083}
  E.~Witten,
  ``Two-dimensional gauge theories revisited,''
  J.\ Geom.\ Phys.\  {\bf 9} (1992) 303
  [arXiv:hep-th/9204083].

\bibitem{9511112}
  A.~S.~Schwarz and O.~Zaboronsky,
  ``Supersymmetry and localization,''
  Commun.\ Math.\ Phys.\  {\bf 183}, 463 (1997)
  [arXiv:hep-th/9511112].

\bibitem{zaboronsky}
O.~Zaboronsky,
``Dimensional reduction in supersymmetric field theories,''
J.\ Phys.\ {\bf A35}, 5511 (2002).

\bibitem{0206161}
  N.~A.~Nekrasov,
  ``Seiberg-Witten Prepotential From Instanton Counting,''
  Adv.\ Theor.\ Math.\ Phys.\  {\bf 7}, 831 (2004)
  [arXiv:hep-th/0206161].

\bibitem{0712.2824}
  V.~Pestun,
  ``Localization of gauge theory on a 
  four-sphere and supersymmetric Wilson
  loops,''
  arXiv:0712.2824 [hep-th].
 
 \bibitem{ati1}
M.F.~Atiyah, Elliptic operators and compact groups. Springer-Verlag,
Berlin,
1974. 

\bibitem{ati2}
P.~Shanahan, The atiyah-singer index theorem : An
introduction, Springer-Verlag.

\bibitem{0608021}
  C.~Beasley, D.~Gaiotto, M.~Guica, L.~Huang, 
  A.~Strominger and X.~Yin,
  ``Why Z(BH) = |Z(top)|**2,''
  arXiv:hep-th/0608021.

\bibitem{0904.4486}
  J.~Gomis, T.~Okuda and D.~Trancanelli,
  ``Quantum 't Hooft operators and 
  S-duality in N=4 super Yang-Mills,''
  arXiv:0904.4486 [hep-th].

\bibitem{0904.4253}
  S.~Murthy and B.~Pioline,
  ``A Farey tale for N=4 dyons,''
  arXiv:0904.4253 [hep-th].

\bibitem{0712.0043}
S.~Banerjee and A.~Sen, 
``Duality Orbits, Dyon Spectrum and Gauge Theory Limit of
Heterotic String Theory on $T^6$'',
arXiv:0712.0043 [hep-th].

\bibitem{0801.0149}
  S.~Banerjee and A.~Sen,
  ``S-duality Action on Discrete T-duality Invariants,''
  arXiv:0801.0149 [hep-th].


\bibitem{0702150}
  A.~Dabholkar, D.~Gaiotto and S.~Nampuri,
  ``Comments on the spectrum of CHL dyons,''
  arXiv:hep-th/0702150.


\bibitem{9507027}
  J.~H.~Schwarz and A.~Sen,
  ``Type IIA Dual Of The Six-Dimensional CHL Compactification,''
  Phys.\ Lett.\  B {\bf 357}, 323 (1995)
  [arXiv:hep-th/9507027].


\bibitem{9607026}
R.~Dijkgraaf, E.~P.~Verlinde and H.~L.~Verlinde,
``Counting dyons in N = 4 string theory,''
Nucl.\ Phys.\ B {\bf 484}, 543 (1997)
[arXiv:hep-th/9607026].

\bibitem{0412287}
G.~L.~Cardoso, B.~de Wit, J.~Kappeli and T.~Mohaupt,
``Asymptotic degeneracy of dyonic N = 4 string states
and black hole
entropy,''
JHEP {\bf 0412}, 075 (2004) [arXiv:hep-th/0412287].

\bibitem{0505094}
  D.~Shih, A.~Strominger and X.~Yin,
  ``Recounting dyons in N = 4 string theory,''
  JHEP {\bf 0610}, 087 (2006)
  [arXiv:hep-th/0505094].

\bibitem{0506249}
D.~Gaiotto,
``Re-recounting dyons in N = 4 string theory,''
arXiv:hep-th/0506249.

\bibitem{0605210}
  J.~R.~David and A.~Sen,
  ``CHL dyons and statistical entropy function from D1-D5 system,''
  JHEP {\bf 0611}, 072 (2006)
  [arXiv:hep-th/0605210].

\bibitem{0802.0544}
  S.~Banerjee, A.~Sen and Y.~K.~Srivastava,
  ``Generalities of Quarter BPS Dyon 
Partition Function and Dyons of Torsion
  Two,''
  arXiv:0802.0544 [hep-th].

\bibitem{0802.1556}
  S.~Banerjee, A.~Sen and Y.~K.~Srivastava,
  ``Partition Functions of Torsion $>1$ Dyons in Heterotic
String Theory on $T^6$,''
  arXiv:0802.1556 [hep-th].

\bibitem{0803.2692}
  A.~Dabholkar, J.~Gomes and S.~Murthy,
  ``Counting all dyons in N =4 string theory,''
  arXiv:0803.2692 [hep-th].

\bibitem{9907200}
  N.~Berkovits, M.~Bershadsky, T.~Hauer, 
  S.~Zhukov and B.~Zwiebach,
  ``Superstring theory on AdS(2) x S(2) as a coset supermanifold,''
  Nucl.\ Phys.\  B {\bf 567}, 61 (2000)
  [arXiv:hep-th/9907200].


\bibitem{0501015}
  A.~Kapustin,
  ``Wilson-'t Hooft operators in four-dimensional gauge theories and
  S-duality,''
  Phys.\ Rev.\  D {\bf 74}, 025005 (2006)
  [arXiv:hep-th/0501015].


\bibitem{romans}
  L.~J.~Romans,
  ``Selfduality For Interacting Fields: 
  Covariant Field Equations For
  Six-Dimensional Chiral Supergravities,''
  Nucl.\ Phys.\  B {\bf 276}, 71 (1986).

\bibitem{9712176}
  F.~Riccioni,
  ``Tensor multiplets in six-dimensional (2,0) supergravity,''
  Phys.\ Lett.\  B {\bf 422}, 126 (1998)
  [arXiv:hep-th/9712176].


      \bibitem{9804166}
        S.~Deger, A.~Kaya, E.~Sezgin and P.~Sundell,
        ``Spectrum of D = 6, N = 4b supergravity on AdS(3) x S(3),''
        Nucl.\ Phys.\  B {\bf 536}, 110 (1998)
        [arXiv:hep-th/9804166].



\end{thebibliography}
\end{document}